\documentclass[10pt,journal,compsoc]{IEEEtran}

\author{Hooman Alavizadeh,~\IEEEmembership{Member,~IEEE,}
	Samin Aref,~
	Dong Seong Kim,~\IEEEmembership{Senior Member,~IEEE,}
	and Julian Jang-Jaccard,~\IEEEmembership{Member,~IEEE}
	
	\IEEEcompsocitemizethanks{\IEEEcompsocthanksitem Hooman Alavizadeh and Julian Jang-Jaccard are with the School of Natural and Computational Sciences, Massey University, New Zealand.\protect\\
		E-mail: h.alavizadeh@massey.ac.nz, j.jang-jaccard@massey.ac.nz
		\IEEEcompsocthanksitem Samin Aref is with the Lab of Digital and Computational Demography at the Max Planck Institute for Demographic Research, Germany\protect\\
		E-mail: aref@demogr.mpg.de
		\IEEEcompsocthanksitem Dong Seong Kim is with the School of Information Technology and Electrical Engineering, University of Queensland, Australia.\protect\\
		E-mail: dan.kim@uq.edu.au
	}
	\thanks{Manuscript received January 04, 2020}}

\usepackage{amssymb}
\usepackage{graphicx}
\usepackage{qtree}
\usepackage{circuitikz}
\usepackage{tikz}
\usepackage{mathtools}
\usepackage{booktabs}
\usepackage{hyperref}
\usepackage{subcaption}
\usepackage{dirtree}
\usepackage{multirow}

\usetikzlibrary{shapes,snakes}

\usepackage{amsthm}

\theoremstyle{definition}
\newtheorem{definition}{Definition}
\newtheorem{example}{Example}

\usepackage[linesnumbered,ruled,vlined]{algorithm2e}
\usepackage{algpseudocode}

\usetikzlibrary{shapes.gates.logic.US,trees,positioning,arrows}

\usepackage{lipsum}

\usetikzlibrary{shapes,arrows}

\usepackage{pgfplots}
\pgfplotsset{width=8.3cm,compat=1.9}

\usepackage{pgfplotstable}
\usepackage{pgfplots}
\usetikzlibrary{patterns}
\pgfplotsset{compat=1.3}
\usepackage{xcolor}
\usepackage{csvsimple}

\usepackage[switch,columnwise]{lineno}
\definecolor{changes}{rgb}{0.8,1,0.95}
\usepackage{xcolor, soul}
\sethlcolor{white}
\usepackage[normalem]{ulem}

\usepackage[utf8]{inputenc}  

\usepackage[linesnumbered,ruled,vlined]{algorithm2e}
\usepackage{algpseudocode}

\makeatletter
\def\@IEEEsectpunct{.\ \,}
\def\paragraph{\@startsection{paragraph}{4}{\z@}{1.5ex plus 1.5ex minus 0.5ex}%
	{0ex}{\normalfont\normalsize\sffamily\bfseries}}
\makeatother

\ifCLASSOPTIONcompsoc
\usepackage[nocompress]{cite}
\else
\usepackage{cite}
\fi
\ifCLASSINFOpdf
\else

\fi

\definecolor{bananamania}{rgb}{0.98, 0.91, 0.71}
\definecolor{babyblueeyes}{rgb}{0.63, 0.79, 0.95}

\begin{document}
	
	
	\title{Evaluating the Security and Economic Effects of Moving Target Defense Techniques on the Cloud}
	
%
	%

\markboth{Evaluating the Security and Economic Effects of MTD Techniques on the Cloud}
{Alavizadeh \MakeLowercase{et al.}: Evaluating the Security and Economic Effects of MTD Techniques on the Cloud}

\IEEEtitleabstractindextext{%
\begin{abstract}

   Moving Target Defense (MTD) is a proactive security mechanism that changes the attack surface with the aim of confusing attackers. Cloud computing leverages MTD techniques to enhance the cloud security posture against cyber threats. While many MTD techniques have been applied to cloud computing, there has so far been no joint evaluation of the effectiveness of MTD techniques with respect to security and economic metrics. In this paper, we first introduce mathematical definitions for the combination of three MTD techniques: Shuffle, Diversity, and Redundancy. Then, we utilize four security metrics – namely, system risk, attack cost, return on attack, and reliability – to assess the effectiveness of the combined MTD techniques applied to large-scale cloud models. Second, we focus on a specific context based on a cloud model for e-health applications to evaluate the effectiveness of the MTD techniques using security and economic metrics. We introduce (1) a strategy to effectively deploy the Shuffle MTD technique using a virtual machine placement technique, and (2) two strategies to deploy the Diversity MTD technique through operating system diversification. As deploying the Diversity technique incurs costs, we formulate the optimal diversity assignment problem (O-DAP), and solve it as a binary linear programming model to obtain the assignment that maximizes the expected net benefit.
\end{abstract}
	
	\begin{IEEEkeywords}
		Cloud Computing, Diversity, Economic Metrics, Redundancy, Security analysis, Shuffle, Optimization.
\end{IEEEkeywords}}

\maketitle

\IEEEdisplaynontitleabstractindextext

%
\IEEEpeerreviewmaketitle

\IEEEraisesectionheading{\section{Introduction}
\label{sec:introduction}}
	
\IEEEPARstart{M}oving Target Defense (MTD) techniques have been proposed that aim to make a system more dynamic, less deterministic, and more unpredictable for cyber attackers by continuously changing the attack surface ~\cite{cho2020toward}. The static nature of systems can make a system more prone to attack, as the attackers have enough time to learn the potential paths of attack, to exploit vulnerabilities, and, ultimately, to penetrate the system. Traditional defensive security solutions for dealing with possible threats, such as anti-malware, intrusion detection systems (IDS), and firewalls, are reactive methods, and tend to be expensive. By contrast, most MTD techniques are proactive defensive methods that adopt the existing technologies in a system (e.g., virtual machines and back-up operating systems (OS)) in order to introduce sufficient levels of unpredictability to confuse the attackers and make security-threatening attempts more complicated. Thus, compared to other approaches, MTD techniques are expected to decrease defensive costs, while increasing the attackers' costs in terms of effort, time, and money. To prevent potential cyber attacks while being economically viable, the proposed defensive MTD strategies must be effective and efficient. In this paper, we leverage MTD capabilities to secure cloud computing. 

We consider the effects of combining different MTD techniques by investigating various MTD properties in situations in which several techniques are deployed. We use the classification of MTD techniques proposed by Hong et al. \cite{hong2016assessing}: Shuffle, Diversity, and Redundancy techniques. While Shuffle MTD techniques are generally designed to enhance the overall security of a system by changing its attack surface, these techniques have no effect on the system's reliability, and may even cause the reliability of the system to deteriorate. The aim of Redundancy MTD techniques is to enhance the reliability or availability of the system. However, because these techniques could place the system in a more vulnerable state, as described in \cite{alavizadeh2021evaluating}, they may actually reduce the overall security of the system (e.g., by increasing the system risk). The aim of Diversity MTD techniques is to make attacks more difficult (e.g., by exploiting the vulnerability of the software). However, as these techniques may increase the cost of defense, they can have negative economic effects. As the outcomes of deploying individual or combined MTD techniques are uncertain, the effectiveness of the proposed MTD techniques has to be evaluated prior to deployment. While using a single MTD technique can be beneficial, problems may arise when trade-offs between security and dependability (such as service availability or reliability) are required. Thus, the question of how MTD techniques can be combined to optimally meet multiple objectives, such as maximizing benefits while reducing undesirable effects, merits extensive investigation.

In this paper, we aim to address the aforementioned problems by evaluating the effectiveness of deploying different MTD techniques, including Shuffle, Diversity, Redundancy; both individually and in combination for cloud computing. Accordingly, we model and analyze MTD techniques using a graphical security model called the hierarchical attack representation model (HARM)~\cite{hong2017towards}. We identify applicable MTD techniques in cloud computing environments, and formally define them. In addition, we use important measures, such as network centrality measures (NCMs) to improve the scalability of the evaluation process for large-sized cloud computing systems.

We investigate the effects of combining different MTD techniques from the three categories, and evaluate using both security and economic metrics by conducting experiments based on two scenarios.

First, we perform an experimental analysis to evaluate how the combined MTD techniques affect the security of the cloud systems from both the attacker's and the defender's perspective. We conduct our experiments using simulation on a large cloud-band model to evaluate the effectiveness of the combined MTD techniques. To examine the level of security these techniques provide from the cloud provider's point of view, we use security metrics, including the system risk and the attack cost. Then, we use the return on attack to evaluate the effectiveness of the combined MTD techniques from the attacker's perspective. Finally, we use other metrics to evaluate the reliability of the cloud after deploying the MTD techniques.

Second, we focus on a more specific context by studying the effectiveness of MTD techniques on economic metric when applied to an e-health cloud model as a case study. To estimate the economic benefits of deploying MTD techniques in this case, we use economic metrics, including the return on security investment and the expected net benefit of security. We utilize both security and economic metrics to show the effectiveness of the proposed MTD techniques. We propose a potentially effective Shuffle strategy, and deploy this technique with the goal of reducing the economic impacts while increasing the security level. We also extend our study by conducting in-depth investigations of Diversity MTD techniques in which we consider the interplay between the costs and the benefits of security. For cloud providers, who face pressure to deploy a defensive strategy on a limited (allocated) budget, using Diversity MTD techniques may be expensive, as they are required to purchase the license and cover the costs of the components' variants (such as VMs). Thus, the use of the components of various systems (such as back-up OS variants) should be precisely prioritized, and possibly optimized. To this end, we propose a Diversity technique strategy based on the globally optimal solution of using an optimization model that maximizes the expected net benefits under all possible Diversity technique assignments.

This paper represents a continuation of a line of studies started in~\cite{alavizadeh2018comprehensive}. In this paper, we extend our contributions as follows: 

\begin{itemize}
    \item We include formalism and the definitions of combined MTD techniques. Using simulation, we evaluate the MTD techniques based on economic and security metrics.
    \item We propose an optimization model that seeks to find an optimal solution to diversity assignment by considering both the costs and the benefits of security.
	\item We provide the formal mathematical definitions for combining the Shuffle, Diversity, and Redundancy MTD techniques.
	\item We evaluate the effectiveness of combined MTD techniques through simulation using security metrics, including system risk ($Risk$), attack cost ($AC$), return on attack ($RoA$), and reliability, for a large cloud model. We evaluate the combined method by deploying the Diversity technique on multiple VMs in the cloud using the OS diversification technique.
	\item We model an e-health cloud example (also called the personal health cloud (PHC)), and evaluate the effectiveness of MTD techniques based on both security and economic metrics.
	\item We provide a set of strategies in which the Shuffle and Diversity techniques can be effectively deployed. We propose a VM placement strategy for the Shuffle technique, and two strategies for deploying the Diversity technique, based on deploying the Diversity technique (OS diversification) (i) with only one back-up OS, and (ii) with multiple back-up OS variants over the set of VMs. 
	\item To solve the second case mentioned above, we propose the optimal diversity assignment problem (O-DAP), and formulate it as a binary linear programming model. This approach allows us to find an assignment of OS variants on multiple VMs while maximizing the expected net benefits.
\end{itemize}

The rest of the paper is organized as follows. In Section~\ref{RV}, we provide a comprehensive overview of the related work and study the MTD techniques. Section~\ref{PR} presents the preliminaries of the paper, including formalisms for the combination of MTD techniques. The proposed MTD definitions and evaluation criteria based on the security metrics are provided in Section~\ref{MTD}. In Section~\ref{Economic}, we evaluate the effectiveness of the MTD techniques using both economic and security metrics as well, and we formulate an optimization model to solve the O-DAP. In Section~\ref{DL}, we continue our discussion, and outline the limitations of the paper. Finally, the paper concludes in Section~\ref{Conclude}.

\section{Related Work}
\label{RV}

\begin{table*}[t]
	\centering
	\caption{MTD techniques applicable in different cloud computing layers}
	\label{layers}
	\begin{tabular}{@{}p{4cm}p{4.5cm}p{3.5cm}p{3.5cm}@{}}
		\toprule
		\multicolumn{1}{l}{Cloud Layer}                                     & Diversity                                                                            & Redundancy                                                   & Shuffle                                                                \\ \midrule
		\begin{tabular}[c]{@{}c@{}}Application Layer\end{tabular}    & \begin{tabular}[c]{@{}l@{}}Web Services \cite{huang2010security, taguinod2015toward}\\ Web Applications \cite{huang2010security, azab2011chameleonsoft}\end{tabular}                      & Web \cite{gorbenko2009using, yuan2013architecture}                                                          & \begin{tabular}[c]{@{}l@{}}Port/IP \cite{antonatos2007defending,luo2014effectiveness,kampanakis2014sdn}\\ Web App. \cite{okhravi2011creating}\\ HTTP \cite{vikram2013nomad, jia2013motag}\end{tabular}      \\
		\begin{tabular}[c]{@{}c@{}}Platform Layer\end{tabular}       & \begin{tabular}[c]{@{}l@{}}Application/ Web service\\ Design \cite{huang2010security}\\ Database \cite{taguinod2015toward}\end{tabular} & Web Server Replica                                           & Web Service                                                            \\
		\begin{tabular}[c]{@{}c@{}}Infrastructure Layer\end{tabular} & \begin{tabular}[c]{@{}l@{}}Operating System (OS) \\ Virtualization \cite{hong2016assessing, alavizadeh2021evaluating,huang2010security}\end{tabular}       & \begin{tabular}[c]{@{}l@{}}Virtualization \cite{hong2016assessing, alavizadeh2021evaluating}\\ SDN \cite{scott2013sdn} \end{tabular} & \begin{tabular}[c]{@{}l@{}}SDN, VM migration \\ \cite{hong2016assessing,9343109, alavizadeh2021evaluating,penner2017combating}\\ Virtual IP \cite{Jafarian:AddMutation2015}\end{tabular} \\ \bottomrule
	\end{tabular}
\end{table*}

A number of studies on MTD theory, techniques, and evaluation have been conducted \cite{cai2016moving,cho2020toward,blakely2019moving}. According to Hobson et al. \cite{hobson2014challenges}, the three main challenges of developing MTD techniques are the coverage, the unpredictability, and the timeline. Zhuang et al. \cite{zhuang2014towards} argued that an effective MTD technique should consider the following issues: (1) which pieces should be moved, (2) whether there is sufficient space for movement, and (3) what the correct time for movement is. Similarly, Cai et al.~\cite{cai2016moving} defined three considerations for the movement of MTD techniques: (1) WHAT to move, (2) HOW to move, and (3) WHEN to move. However, these studies have not discussed the cost or economic efficiency of the MTD movement. Rather, they merely explored the common properties of the MTD techniques (movement selection, movement strategy, and movement time) that should be realized when an MTD technique is adopted. Thus, in addition to considering the total cost of security, providers should take into account the cost of movement relative to the level of security achieved.

Our analysis framework contributes to the literature by including (1) MTD techniques and categories, (2) MTD applicable layers, and (3) a definition of the combination of MTD techniques at different layers of the cloud~\cite{cho2020toward} (as shown in Table \ref{layers}). 

Hong et al. \cite{hong2016assessing} classified the MTD strategies according to three comprehensive categories as follows. The Shuffle technique refers to any rearrangement of the system setting into different software, hardware, and network layers, like changing or shuffling the IP address; rearranging the network's topology; or moving or migrating a VM, a host, or hardware to another location \cite{penner2017combating,alavizadeh2019automated}. The Diversity technique involves replacing the components' variant, such as a server, a programming language, an operating system, or hardware, while the system continues to provide functionality equivalent to that of the previous state (before changing the variant)~\cite{hong2016assessing}. The Redundancy technique can be used increase the number of replica components in the system, such as servers, hardware, OS, software, and services~\cite{yuan2013architecture}. 

The Shuffle technique has been proposed in several studies. Most of the existing research focused on the novelty of the implementation and the application of the Shuffle technique.
Jafarian et al.~\cite{Jafarian:AddMutation2015} implemented an IP shuffling technique that mutates IP addresses unpredictably. They focused on the minimization of the overhead of this operation after each IP mutation. 
Moreover, the application of Shuffle techniques on the cloud has been studied in~\cite{danev2011enabling, alavizadeh2021evaluating}. Danev et al. \cite{danev2011enabling} proposed a Shuffle technique for securing the cloud infrastructure. They focused on secure VM migration in the cloud. Their approach was to utilize an extra physical trusted platform module and trusted parties for the migration process. They also used public key infrastructure to secure the protocol. In addition, they suggested a comprehensive evaluation of different criteria, like assessing the main security services (CIA triad), and analyzing how the migration scenario performs in terms of time and RAM size usage against cryptography protocols. Penner and Guirguis in \cite{penner2017combating} developed a set of MTD technologies to change the location of VMs in the cloud in order to defend against MultiArmed Bandit (MAB) attacks caused by weak VM isolation in the cloud. They actually deployed a MTD technique based on the attacker's point of view, and showed that their method can thwart a MAB attack designed to find critical information (e.g., databases and credit card information). They assessed the performance of the proposed method by measuring the time of the switch in VMs. However, most of the existing techniques focused only on minimizing the overhead or improving the performance, and the security-related impacts of deploying MTD techniques have not been evaluated using security models. Therefore, there is a lack of research on the economic as well as the security impacts of deploying MTD techniques on the cloud.
	
A method for deploying the Diversity technique in the programming language has been proposed by Taguinod et al. in~\cite{taguinod2015toward}. Moreover, a Diversity technique for deployment on virtual servers has been proposed by Huang et al. \cite{huang2010security} with the aim of improving the resiliency of the network and services. They have evaluated their method by computing the probability of attack success. In~\cite{azab2011chameleonsoft}, the authors developed a Diversity technique designed to change a running program's variants erratically to enable a large program to be divided into smaller components (tasks). They used a recovery mechanism to enhance the system's resilience. The idea is to use a different variant at runtime to confuse the attacker. The application of the Diversity technique on the cloud has been investigated in~\cite{alavizadeh2018evaluation, alavizadeh2020model,hong2016assessing} from a security perspective. However, these studies did not consider the economic impacts of deploying the Diversity technique on the cloud. Redundancy techniques have been proposed in~\cite{gorbenko2009using, yuan2013architecture}. A Redundancy technique for use on the application layer has been introduced by Gorbenko et al. in \cite{gorbenko2009using}. They proposed a method for web service replication designed to improve the dependability of the system. They evaluated their method through performance analyses like assessments of system response time and availability. Another Redundancy technique has been proposed by Yuan et al. in~\cite{yuan2013architecture}. They proposed an approach for deploying the Redundancy technique on web services that aims to prevent malicious code injection attacks on the servers. However, they did not assess the effectiveness of their proposed method through security analysis.

There is a gap in the MTD research due to the lack of evaluations of the proposed MTD techniques based on security and economic metrics. Only a few studies, such as ~\cite{bistarelli2012evaluation}, have proposed evaluating the economic impacts of defensive techniques using graphical security models (GSMs) and economic metrics. Security metrics can be incorporated into GSMs to evaluate the effectiveness of both the given network models and the MTD techniques.

\section{Definitions and Formalization}\label{PR}
\subsection{E-Health Cloud Model}
\begin{figure}[t]
	\centering
	\includegraphics[width=0.48\textwidth]{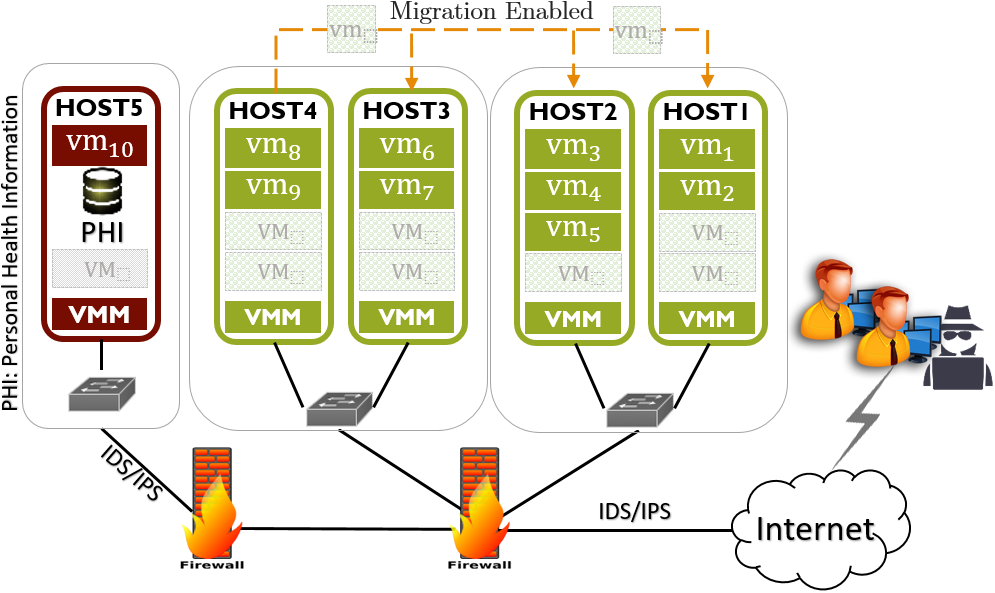}
	\caption{an E-Health cloud model including PHI records}
	\label{fig:Cloud-Model}
\end{figure}

{We consider a private personal health cloud (PHC) that includes 10 VMs located on different cloud hosts (servers). We assume that VMs $vm_1$ and $vm_2$ in Host1 are connected to the internet (entry points of the cloud), and that the last VM $vm_{10}$ in Host5 is connected to a critical database (DB) that includes the personal health information (PHI) of patients, as shown in Figure~{\ref{fig:Cloud-Model}}. The VMs located in Host1 and Host2 use Windows 10, and the VMs in the other hosts are installed with Enterprise Linux OS. We assume that an attacker is outside the private cloud and can exploit the vulnerabilities of those operating systems to gain access. The goal of the attacker is to compromise the database (DB) in Host5.
Table~{\ref{vul}} shows the vulnerabilities for different OSs based on the National Vulnerability Database (NVD)} \cite{mell2006common}. {We assume that there are three vulnerabilities for both Windows OS and Linux OS, and that there is a single vulnerability for Fedora OS.}

\begin{figure}[t]
	\centering
	\begin{tikzpicture}[scale=0.65, every node/.style={transform shape}]
	\node[shape=circle,draw=black,align=center,fill=red!10]  (A) at (9,0) {$A$};
	\node[shape=rectangle,line width=0.1pt, rounded corners,opacity=.8,draw=black,align=center,fill=red!20] (W) at (9+0.90,0+0.50) {\footnotesize Outside of cloud};
	\node[shape=circle,draw=black,align=center] (v1) at (7,2) {$\text{vm}_1$};
	\node[shape=rectangle,line width=0.1pt, rounded corners,opacity=.8,draw=black,align=center,fill=green!35] (W) at (7+0.45,2+0.50) {\footnotesize$W$};
	\node[shape=circle,draw=black,align=center] (v2) at (7.7,-2) {$\text{vm}_2$};
	\node[shape=rectangle,line width=0.1pt, rounded corners,opacity=.8,draw=black,align=center,fill=green!35] (W) at (7.7+0.45,-2-0.50) {\footnotesize$W$};
	\node[shape=circle,draw=black,align=center] (v4) at (6,0) {$\text{vm}_4$};
	\node[shape=rectangle,line width=0.1pt, rounded corners,opacity=.8,draw=black,align=center,fill=green!35] (W) at (6+0.65,0) {\footnotesize$W$};
	\node[shape=circle,draw=black,align=center] (v5) at (5.5,-2.3) {$\text{vm}_5$};
	\node[shape=rectangle,line width=0.1pt, rounded corners,opacity=.8,draw=black,align=center,fill=green!35] (W) at (5.5+0.45,-2.3+0.50) {\footnotesize$W$};
	\node[shape=circle,draw=black,align=center] (v3) at (4.4,2.4) {$\text{vm}_3$};
	\node[shape=rectangle,line width=0.1pt, rounded corners,opacity=.9,draw=black,align=center,fill=green!35] (W) at (4.4+0.4,2.4+0.45) {\footnotesize$W$};
	\node[shape=circle,draw=black,align=center] (v7) at (4.2,-1.1) {$\text{vm}_7$};
	\node[shape=rectangle,line width=0.1pt, rounded corners,opacity=.9,draw=black,align=center,fill=orange!55] (W) at (4.2+0.4,-1.1+0.45) {\footnotesize$L$};
	\node[shape=circle,draw=black,align=center] (v6) at (2.8,1.1) {$\text{vm}_6$};
	\node[shape=rectangle,line width=0.1pt, rounded corners,opacity=.9,draw=black,align=center,fill=orange!55] (W) at (2.8-0.4,1.1+0.45) {\footnotesize$L$};
	\node[shape=circle,draw=black,align=center] (v9) at (2.1,-1.8) {$\text{vm}_9$};
	\node[shape=rectangle,line width=0.1pt, rounded corners,opacity=.9,draw=black,align=center,fill=orange!55] (W) at (2.1+0.4,-1.8-0.45) {\footnotesize$L$};
	\node[shape=circle,draw=black,align=center] (v8) at (0.85,1.8) {$\text{vm}_8$};
	\node[shape=rectangle,line width=0.1pt, rounded corners,opacity=.9,draw=black,align=center,fill=orange!55] (W) at (0.85+0.4,1.8+0.45) {\footnotesize$L$};
	\node[shape=circle,draw=black,align=center,fill=red!10] (v10) at (-0.5,-0.2) {$\text{vm}_{10}$};

	\path [dashed,line width=0.9pt,red,->] (A) edge node[left] {} (v1);
	\path [->] (A) edge node[left] {} (v2);
	\path [->] (v1) edge node[left] {} (v3);
	\path [dashed,line width=0.9pt,red,->] (v1) edge node[left] {} (v4);
	
	\path [->] (v2) edge node[left] {} (v4);
	\path [->] (v2) edge node[left] {} (v5);
	
	\path [->] (v3) edge node[left] {} (v5);
	\path [->] (v3) edge node[left] {} (v6);
	
	\path [->] (v4) edge node[left] {} (v5);
	\path [dashed,line width=0.9pt,red,->] (v4) edge node[left] {} (v6);
	
	\path [->] (v5) edge node[left] {} (v7);
	\path [->] (v5) edge node[left] {} (v9);
	
	\path [->] (v6) edge node[left] {} (v8);
	\path [dashed,line width=0.9pt,red,->] (v6) edge node[left] {} (v9);
	
	\path [->] (v7) edge node[left] {} (v6);
	\path [->] (v7) edge node[left] {} (v9);
	
	\path [->] (v8) edge node[left] {} (v10);	
	\path [dashed,line width=0.9pt,red,->] (v9) edge node[left] {} (v10);
	
	\tikzset{grow'=down}
	\tikzset{every tree node/.style={anchor=base west}}
	\tikzstyle{level 1}=[sibling distance=7mm]
	\node[shape=or gate US,thick,rotate=90,draw=black] (at8) at (1.8,-3.9) {OR}
	child
	{
		child
		{
			node{\large{$\nu_{3,\textsc{l}}$}}
		}
		child
		{
			node{\large{$\nu_{2,\textsc{l}}$}}
		}
		child
		{
			node{\large{$\nu_{1,\textsc{l}}$}}
		}
	};	
	\node[] (dots) at (3,-3.9) {$\dots$};
	\tikzset{grow'=down}
	\tikzset{every tree node/.style={anchor=base west}}
	\tikzstyle{level 1}=[sibling distance=7mm]
	\node[shape=or gate US,thick,rotate=90,draw=black] (at6) at (4.2,-3.9) {OR}
	child
	{
		child
		{
			node{\large{$\nu_{3,\textsc{l}}$}}
		}
		child
		{
			node{\large{$\nu_{2,\textsc{l}}$}}
		}
		child
		{
			node{\large{$\nu_{1,\textsc{l}}$}}
		}
	};	
	
	\tikzset{grow'=down}
	\tikzset{every tree node/.style={anchor=base west}}
	\tikzstyle{level 1}=[sibling distance=9mm]
	\node[shape=or gate US,thick,rotate=90,draw=black] (at10) at (-0.5,-3.9) {OR} 
	child
	{
		child
		{
			node{\large{$\nu_{3,\textsc{l}}$}}
		}
		child
		{
			node{\large{$\nu_{2,\textsc{l}}$}}
		}
		child
		{
			node{\large{$\nu_{1,\textsc{l}}$}}
		}
	};
	\node[shape=or gate US,thick,rotate=90,draw=black] (at1) at (7,-3.9) {OR}
	child
	{
		child
		{
			node{\large{$\nu_{3,\textsc{w}}$}}
		}
		child
		{
			node{\large{$\nu_{2,\textsc{w}}$}}
		}
		child
		{
			node{\large{$\nu_{1,\textsc{w}}$}}
		}
	};
	
	\path [-] [dashed] (v10) edge node[left] {} (at10);
	\path [-] [dashed] (v8) edge node[left] {} (at8);
	\path [-] [dashed] (v1) edge node[left] {} (at1);
	\path [-] [dashed] (v7) edge node[left] {} (at6);		
	\end{tikzpicture}
	\caption{Generated two-layer HARM for the cloud. An example of an attack path is highlighted as a dashed line.} 
	\label{fig:HARM1}
\end{figure}
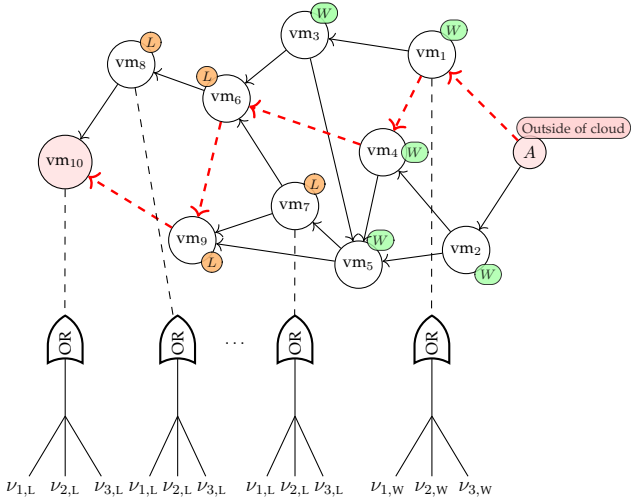

\subsection{HARM Construction}
\label{Def-HARM}
In order to perform the security analysis of the cloud, and to further evaluate the effects of MTD techniques on the cloud, we used the two-layered HARM~\cite{hong2017towards} to model the cloud. The HARM consists of an upper layer to model the connectivity of VMs, and a lower layer to capture the vulnerabilities on each VM. Using the HARM, we can compute security metrics for the purposes of comparison. We evaluate how the security metrics change when the MTD techniques are deployed later, in Section~\ref{MTD}.

\theoremstyle{definition}
\begin{definition}\label{HARM}
	{We can show a HARM}~\cite{hong2012harms} {as a 3-tuple $H=(U,L,C)$ where $U$ refers to the upper layer, which is an attack graph (AG) and $L=\{\ell_1,\ell_2,\dots,\ell_n\}$ is a set of attack trees (ATs) in the lower layer. We define $C: U \rightarrow L$ as a one-to-one mapping of the upper layer to the lower layer. The relation $C$ relates the graph $U$ to the set $L$. The upper layer of the HARM is defined as a graph $U = (\text{VM},\text{E})$, where $\text{VM} = \{vm_1,vm_2,\dots,vm_n\}$ is a set of VMs in the cloud and $E \subset VM \times VM$ is a set of connectivities between the VMs denoted as edges. An edge can be a communication medium between the VMs $vm_s, vm_t\in VM$ when an exploit can be performed. The member $\ell_i$ of the set $L$ is an AT corresponding to $vm_i$ in the upper layer. The nodes of the tree $\ell_i$ are $\{V_{i,\theta},G,root_i\}$ including (1) the set of vulnerabilities $V_{i,\theta}$ corresponding to VM index $i$ and operating system $\theta$, (2) the function $G$, and (3) $root_i$ which is the virtual machine index $i$ in the upper layer.
	The set $V_{i,\theta}$ of vulnerabilities is defined as $V_{i,\theta}=\{\nu_{1,\theta} ,\nu_{2,\theta},\dots,\nu_{m,\theta}\}$. The set of operating systems is $OS=\{W,L,F\}$ (including \textbf{W}indows10, \textbf{L}inux, and \textbf{F}edora), and $\theta$ is a given member of it $\theta \in OS$.
	The function $G$ applies logical gates to $V_{i,\theta}$. The logical gates may be any combination of $AND\text{-gate}, OR\text{-gate}$ depending on how the vulnerabilities might be exploited by the attacker.
	}
\end{definition}

\begin{table}[t]
	\centering
	\caption{{OS Vulnerabilities ($V$) including Base-Score ($BS$), Impact ($I$), Exploitability ($E$), and Attack Cost ($AC$)}}
	\label{vul}
	\begin{tabular}{@{}lllllll@{}}
		\toprule
		OS ($\theta$)                    & $V$ & CVE-ID         & $BS$ & $I$ & $E$ & $AC$ \\ \midrule
		\multirow{3}{*}{\textbf{W}in10} & $\nu_{1,\textsc{w}}$ & CVE-2018-8490  & 8.4   & 6    & 0.17 & 1.6   \\
		& $\nu_{2,\textsc{w}}$ & CVE-2018-8484  & 7.8   & 5.9  & 0.18 & 2.2   \\
		& $\nu_{3,\textsc{w}}$ & CVE-2016-3209  & 8.8   & 5.9  & 0.28 & 1.2   \\ \midrule
		\multirow{3}{*}{\textbf{L}inux} & $\nu_{1,\textsc{l}}$ & CVE-2018-14678 & 7.8   & 5.9  & 0.18 & 2.2   \\
		& $\nu_{2,\textsc{l}}$ & CVE-2018-14633 & 7     & 4.7  & 0.22 & 3     \\
		& $\nu_{3,\textsc{l}}$ & CVE-2017-15126 & 8.1   & 5.9  & 0.22 & 1.9   \\ \midrule
		\textbf{F}edora                 & $\nu_{1,\textsc{f}}$ & CVE-2014-1859  & 5.5   & 3.6  & 0.18 & 4.5   \\ \bottomrule
	\end{tabular}
\end{table}

{Figure~{\ref{fig:HARM1}} illustrates an example of the constructed HARM for the cloud system that includes 10 VMs with different OSs (either Windows or Linux). Note that the upper layer of HARM captures the conectivities of those VMs, while the lower layer shows the vulnerabilities associated with each VM. In this paper, we assume that the attacker needs to exploit only one vulnerability to exploit the VM. Thus, we only use the OR-gate in the lower layer of the HARM. However, a combination of the AND-gate and the OR-gate in the attack tree can be used if more complicated exploit scenarios are needed to exploit a system, as explained in}~\cite{roy2012attack}. {However, in the interests of the readability of our proposed HARM, more complicated vulnerability analyses, such as those discussed in}~\cite{roy2012attack}, {are considered outside the scope of this paper.}

\subsection{Importance Measures}
As we stated earlier, the upper layer of the HARM represents a comprehensive scheme for the connectivity of VMs in the cloud through a graph. To efficiently carry out the security analysis, we need to identify the most important components (here, the VMs) of the network. {Network centrality measures (NCMs) such as betweenness and closeness have been widely used as importance measures (IMs) in security modeling and analysis in various contexts, such as in cloud computing} \cite{alavizadeh2021evaluating} {and enterprises~}\cite{yusuf2016security},{ as well as in other contexts}~\cite{cadini2008using}.{ Thus, we use two important NCMs, closeness and betweenness} \cite{cadini2008using}{, to identify more crucial VMs in the cloud.} In Section~\ref{MTD}, we show how IMs can be utilized to find more effective MTD strategies.
	
\begin{definition}
	NCMs can be computed for the upper layer of the HARM defined in Definition~\eqref{HARM}. 
	We can calculate the closeness centrality ($C_c$) of a specific VM in the network in Equation~\eqref{eq:closeness}.
	
	\begin{equation}\label{eq:closeness}
	C_c(vm_i)= (n-1)\Bigg(\sum_{j\in VM} d(vm_i,vm_j)\Bigg)^{-1}
	\end{equation}
	
	{In Equation~{\eqref{eq:closeness}}, $d(vm_i,vm_s)$ is a function computing the length of the shortest path between VMs $vm_i$ and $vm_s$ in the HARM, and $n$ denotes the total number of VMs in the HARM.} Then, the betweenness centrality ($C_b$) of a VM can be computed using Equation~\eqref{eq:betweenness}.
	
	\begin{equation}\label{eq:betweenness}
	C_b(vm_i)= \sum_{s,t \in VM \setminus \{vm_i\}} \frac{\delta_{st}(vm_i)}{\delta_{st}},
	\end{equation}
	
	In Equation~\eqref{eq:betweenness}, $\delta_{st}$ is a function calculating the total number of the shortest paths between each pair of VMs $(s,t) \in VM$, and $\delta_{st}(vm_i)$ denotes the number of such paths passing through the specific VM ($vm_i$).
\end{definition}

\subsection{Security Metrics}
In this section, we utilize four security metrics, (i) system risk ($Risk$), (ii) attack cost ($AC$), (iii) return on attack ($RoA$), and (iv) system reliability ($Reliability$), to evaluate the security of the cloud after deploying MTD techniques, and to identify the most suitable technique deployment strategies. Risk is based on the vulnerabilities of the network's components~\cite{hong2016assessing}. $AC$ measures the difficulties attackers face in attacking a system, and can be quantified in terms of the costs an attacker incurs when seeking to exploit a network component or the whole system~\cite{yusuf2016security}. $RoA$ indicates the willingness of the attacker to use the same components, attack path(s), and vulnerabilities to penetrate the network. $RoA$ quantifies the costs of an attack versus the benefits of the attack~\cite{cremonini2005evaluating}. $Reliability$ quantifies the reliability of the network's components (i.e., critical components) under certain attack circumstances. The $Reliability$ can be calculated using the symbolic hierarchical automated reliability and performance evaluator (SHARPE) that is an analytic modeling software tool \cite{sahner2012performance}.

\subsubsection{System Risk}
{We define system risk ($Risk$) as the overall risk value associated with the cloud system, which can be computed through the HARM. To compute the $Risk$, the HARM should be constructed by obtaining information related to the connectivities of VMs on the cloud for the upper layer, and by obtaining the vulnerabilities of each VM for the lower layer. Note that the vulnerabilities associated with each VM in the cloud can be obtained through a vulnerability database (i.e., NVD). 
Let $E_{\nu_{j,\theta}}$ and $I_{\nu_{j,\theta}}$ be the exploitability and impact values of $j^{th}$ vulnerability existing on the OS $\theta$ on the $vm_i$ such that $\nu_{j,\theta} \in V_{i,\theta}$. 
Then, the risk of a VM can be computed as the product of those values returning the maximum value, as shown in Equation~{\eqref{eq:impact}}.}
\begin{equation}\label{eq:impact}
Risk_{vm_i}=\max_{\nu_{j,\theta} \in V_{i,\theta}} \big( E_{\nu_{j,\theta}} \times I_{\nu_{j,\theta}} \big)
\end{equation}
{Then, the risk of an attack path $Risk_{ap}$ can be computed by counting all risk values on any single VM in an attack path ($ap$) from the attacker to the target as Equation~{\eqref{eq:Rap}}. Finally, the sum of all risk values associated with all possible attack paths on the cloud provides the $Risk_c$ value. Note that the set of all possible attack path is denoted as $AP$, see Equation~{\eqref{eq:Rc}}.}
\begin{equation}\label{eq:Rap}
Risk_{ap} = \sum_{vm_i \in ap} Risk_{vm_i}
\end{equation}

\begin{equation}\label{eq:Rc}
Risk_c = \sum_{ap \in AP} Risk_{ap}
\end{equation}

\begin{example}\label{CH4:ap}
	Figure~{\ref{fig:HARM1}} shows an example of an attack path ($ap_1$) from the attacker to the target. The risk value associated with this attack path can be computed as $Risk_{ap_1}=Risk_{vm_1}+Risk_{vm_4}+Risk_{vm_6}+Risk_{vm_9}+Risk_{vm_{10}}= 5.9*0.28+ 5.9*0.28+ 5.9*0.22+5.9*0.22$ $+ 5.9*0.22=\textbf{7.2}$. Finally, the total risk of the cloud is the sum of all $Risk_{ap}$ values which is $Risk_c =\textbf{211.692}$ for the cloud example.
\end{example}

\subsubsection{Attack Cost}
{The cost of exploiting the vulnerabilities on a VM for an attacker is defined as the attack cost ($AC$). We expand this metric to compute the overall $AC$ of the cloud. We use the upper layer of the HARM to compute the $AC$. Table~{\ref{vul}} lists the costs of exploiting a VM through vulnerabilities ($AC_{vm}$). Equation~{\eqref{eq:ACs}} shows the formula for the overall $AC$ value of the cloud.}

\begin{equation}\label{eq:ACs}
AC_c = \sum_{ap \in AP} \bigg( \sum_{vm_i \in ap} AC_{vm_i} \bigg)
\end{equation}
{where $ap$ is a single attack path in the system and $AP$ is the list of all possible attack paths in the network.}

\subsubsection{Return on Attack}
{Return on attack ($RoA$) is a security metric based on the attacker's perspective~}\cite{cremonini2005evaluating}. {$RoA$ quantifies the costs of the attack versus the benefits of the attack. A higher value of $RoA$ indicates a higher probability that an attacker will exploit those vulnerabilities (a higher tendency to attack). The ratio of the risk value for a VM and the attack cost determines the $RoA$ value for a specific VM, which is shown in Equation {\eqref{CH4:eq:RoAvm}}. Then, the overall $RoA$ of a system can be computed through Equation~{\eqref{CH4:eq:RoAs}}.}  

\begin{equation}\label{CH4:eq:RoAvm}
RoA_{vm_i} = \frac{\max_{\nu_{j,\theta} \in V_{i,\theta}} \big( E_{\nu_{j,\theta}} \times I_{\nu_{j,\theta}} \big)}{AC_{vm_i}}
\end{equation}

\begin{equation}\label{CH4:eq:RoAs}
RoA_{c} =\sum_{ap \in AP}\bigg(\sum_{vm_i \in ap} RoA_{vm_i}\bigg)
\end{equation}

\begin{figure*}[t]
	\centering
	\begin{subfigure}[b]{0.42\textwidth}
		\centering
		\includegraphics[height=7.5cm]{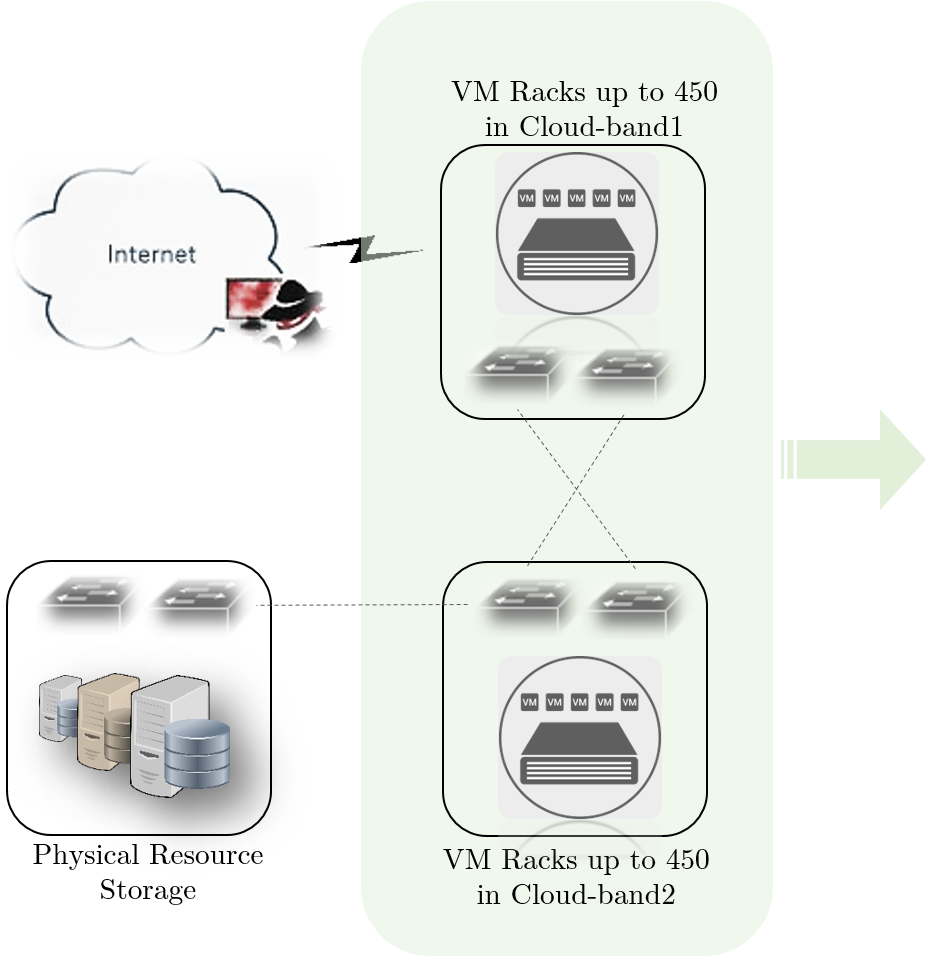}
		\caption{}
		\label{fig:cloud}
	\end{subfigure}
	\begin{subfigure}[b]{0.57\textwidth}
		\includegraphics[height=7.8cm]{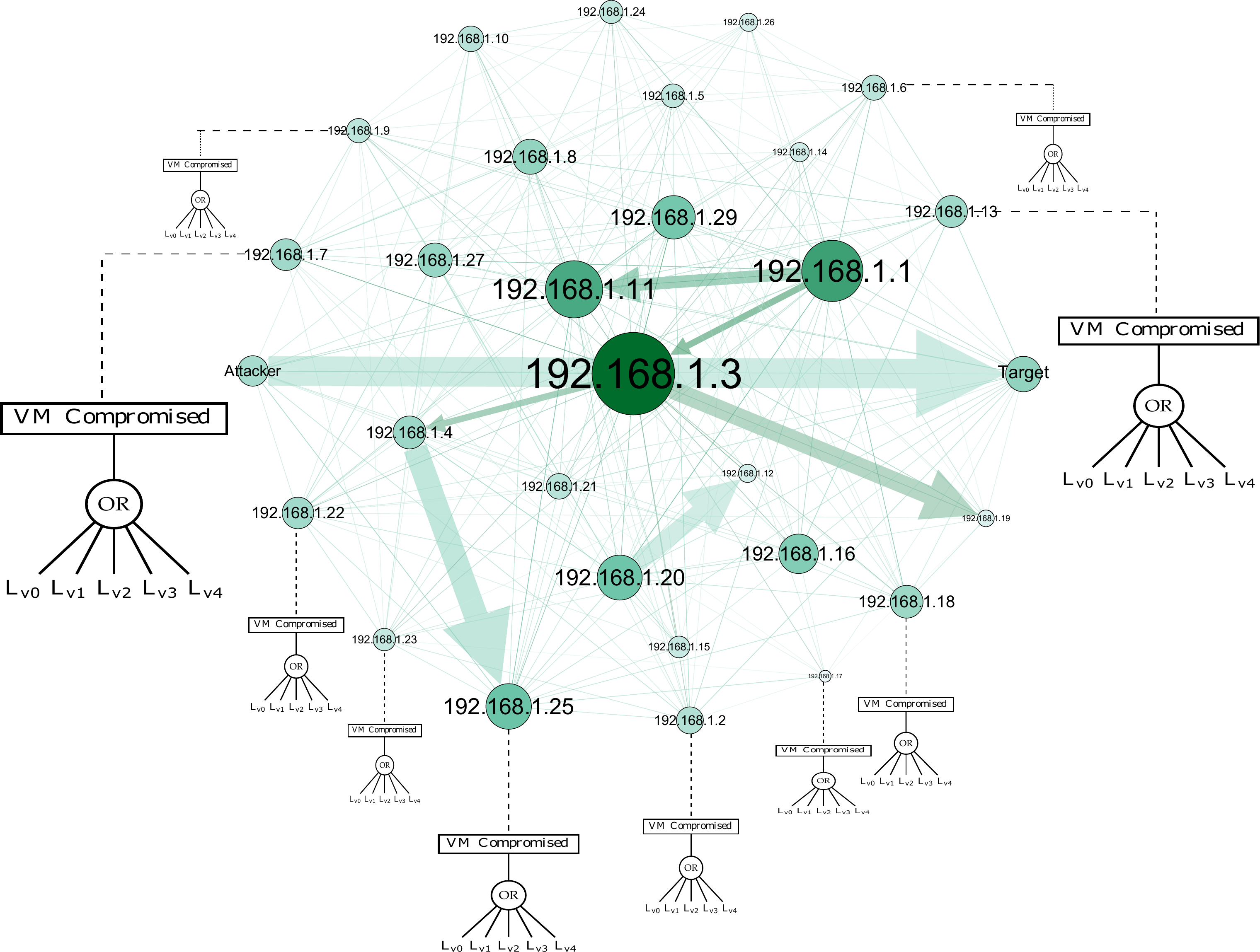}
		\caption{}
		\label{fig:HARM}
	\end{subfigure}
	\caption{The Cloud-band model. (a) The Cloud-band model, including two cloud-band nodes and one resource node. (b) An example of the generated HARM for the cloud-band model with 30 VMs, which shows the VM's connectivities and vulnerabilities in the upper and the lower layers, respectively (note that this cloud-band model can hold up to 450 VMs).}
	\label{fig:cloud-band}
\end{figure*}
	
\subsection{MTD Formalism}		
We utilize the virtual machine live migration (VM-LM) as the main technique for deploying the Shuffle MTD technique, which can be formulated based on the HARM definition as follows.

\begin{definition}\label{SD}
	Let $S(H,\kappa)$ be a Shuffle function on the HARM where $1\leq \kappa \leq n$, and $\kappa$ denotes a specific VM $vm_{\kappa} \in VM$ chosen for the shuffling procedure. Then, the result of the Shuffle function is as $S(H,\kappa)=H^\text{s}$. We define $H^\text{s}=(U^{\text{s},\kappa},L,C)$ where $U^{\text{s},\kappa}$ is the transformed AG resulting from the Shuffle on $vm_k$ in the upper layer of the HARM, and can be represented as $U^{\text{s},\kappa} = (VM,E')$.
\end{definition}

The Diversity technique is formulated as follows.
\begin{definition}\label{DD}
	We formulate the Diversity technique in which the Diversity function is applied to $H$ as $D(H,\kappa)=H^d$, where $\kappa$ denotes a specific VM $vm_{\kappa} \in VM$ selected for being replaced with another OS variant. Then, $H^d=(U,L^{d,k},C)$ is the result of deploying the Diversity technique, where $L^{d,k}=\{\ell_1,\dots,\ell_\textbf{k},\dots,\ell_n\}$ denotes the ATs corresponding to each VM and $\ell_k=(V_{k,\theta},G,root)$ is the transformed AT of $vm_k$ which is replaced with another variant $\theta \in OS$. The Diversity technique affects the lower layer and varies vulnerabilities $V_{k,\theta}$$=$$\{\nu_{1,\theta}, \nu_{2,\theta},\dots, \nu_{m,\theta}\}$, while $U$$=$$(VM,E)$ is preserved. 
\end{definition}

We formulate the Redundancy technique as follows.
\begin{definition}
	Let $R(H,k,r)$ be a Redundancy function on the HARM where $k$ denotes the VM that should be replicated $r$ times. Then, the resulting Redundancy function is $R(H,k,r)=H^\text{r}$, where $1\leq k \leq n$ and $r \leq l$, and $l$ is a limit for the replication of a VM. Thus, $vm^\text{r}_k$ shows the replicated VM in the upper layer. We define $H^\text{r}=(U^\text{r}_k,L^\text{r}_k,C)$ where $U^\text{r}_k$ is a transformed AG resulting from $r$ times replication of $vm_k$ in the upper layer of the HARM, and can be represented as $U_{k}^\text{r} = (VM',E')$, where $VM'$ can be shown as: $$VM'=VM\cup \bigg(\bigcup_{r=1}^{l} vm^\text{r}_k\bigg)$$ and $|VM'| = n+r$, and $E^\prime \subset VM' \times VM'$.
\end{definition}
The replication of VMs in the upper layer results in adding vulnerabilities to the lower layer of the HARM. We can then define the lower layer as follows.
\begin{definition}
	Suppose that $V^\text{r}$ is a set of vulnerabilities caused by the replication of a VM in the upper layer of the HARM. {The lower layer of the HARM can then be updated based on $L_{k}^\text{r}$ including $\{V',G,root\}$, where $V'=V\cup V^\text{r}$ is a set of new vulnerabilities introduced by adding the new VMs.}
\end{definition}
We formulate the combination of Shuffle, Diversity, and Redundancy (S+D+R) as a function on the HARM as follows.
\begin{definition}
	Let $\text{S+D+R}(H,k_s,k_d,k_r,r)$ be a S+D+R function on the HARM where $k_r$ shows the VM selected to be replicated $r$ times, $k_s$ is the VM selected to be shuffled, and $k_d$ denotes the VM selected for the Diversity technique. Then, the resulting S+D+R function is $\text{S+D+R}(H,k_s,k_d,k_r,r)=H^{\text{s+d+r}}$, where $1\leq k_s,k_d,k_r \leq n$ and $0<r\leq l$.
	We define $H^\text{s+d+r}=(U^{\text{s+d+r}}_{k_r,k_s},L^{\text{s+d+r}}_{k_r,k_d},C)$ where $U^{\text{s+r}}_{k_r,k_s}$ is a transformed AG in the upper layer in which S+R is deployed on and $L^{\text{s+r}}_{k_r,k_d}$ is the corresponding transferred AT in the lower layer. Then, the former can be represented as $U^{\text{s+d+r}}_{k_r,k_s} = (VM',E')$, where $VM'$ can be shown as: $$VM'=VM\cup \bigg(\bigcup_{r=1}^{l} vm^\text{r}_{k_r}\bigg)$$ and $|VM'| = n+r$, and $E^\prime \subset VM' \times VM'$. Next, the latter can be shown as $L_{k_r,k_d}^\text{s+d+r} = (V',G,root)$, where $V'=V\cup V^r \cup V^d$, and $V^\text{r}$ is a set of vulnerabilities caused by the replication of a VM, and $V^\text{d}$ is a set of new vulnerabilities introduced by replacing the OS.
\end{definition}



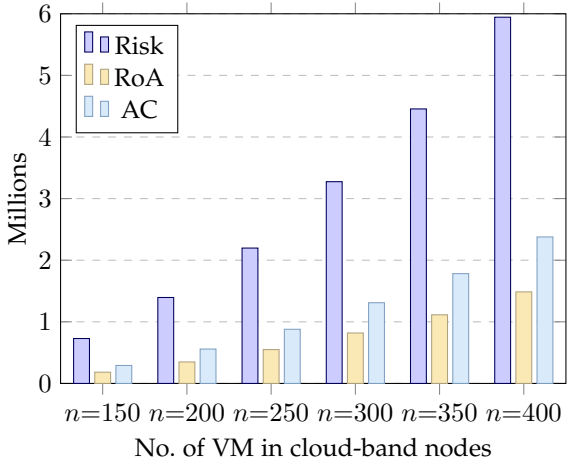
\begin{figure}[t!]
	\centering
	\begin{tikzpicture}
	\begin{axis}[
	ybar,
	height=6.5cm,
	width=8.3cm,
	bar width=6pt,
	y label style={at={(axis description cs:-0.05,.5)},anchor=south},
	xlabel={No. of VM in cloud-band nodes},
	ylabel={Millions},
	xmin=0.5, xmax=6.5,
	ymin=0, ymax=6,
	xtick={0,1,2,3,4,5,6},
	ytick={0,1,2,3,4,5,6},
	xticklabels={,$n$$=$$150$,$n$$=$$200$,$n$$=$$250$,$n$$=$$300$,$n$$=$$350$,$n$$=$$400$},
	legend pos=north west,
	ymajorgrids=true,
	grid style=dashed,
	]
	
	\addplot[
	color=blue!35!black,fill=blue!20!white,
	]
	table [x=VM, y=Risk, col sep=comma]{data.csv};

	\addplot[
	color=bananamania!70!black,fill=bananamania,
	]
	table [x=VM, y=RoA, col sep=comma]{data.csv};	
	
	\addplot[
	color=babyblueeyes!80!black,fill=babyblueeyes!40!white,
	]
	table [x=VM, y=Cost, col sep=comma]{data.csv};	
	
	\legend{$\text{Risk}$\\
		$\text{RoA}$\\
		$\text{AC}$\\},
	\end{axis}
	\end{tikzpicture}
	\caption{Security analysis results of the current cloud-band}
	\label{fig:current-metrics}
\end{figure}

\section{MTD Security Analysis}\label{MTD}
\subsection{Large Cloud Model}
In this section, we use an example of a large cloud model based on the cloud-band system. A cloud-band system typically consists of two main cloud-band nodes, with each accommodating up to 450 VMs, and one resource node connecting to a database (DB). The (cloud) system model can be used as an input to generate the graphical security models, with HARM formalism described in Subsection~\ref{Def-HARM}, which can be used for evaluating the security of the cloud. Figure \ref{fig:cloud-band} demonstrates the abstract cloud-band model, and {Figure~{\ref{fig:HARM}} shows the generated HARM of the cloud-band. Note that the VMs are connected in a mesh topology. The generated HARM for 400VMs has the following properties. the mode of path lengths are 10.53 and eight, respectively, with the shortest path between the attacker and the DB being eight. The standard deviation of the attack path lengths is 1.36. We compute the betweenness of all VMs. The highest values for the three most important VMs in the cloud-band are 0.35, 0.27, 0.21, respectively. The minimum betweenness value in the upper layer of the HARM is zero.} In this model, we assume that an attacker is outside of the cloud-band, and can penetrate the cloud by exploiting the vulnerabilities of VMs in the first cloud band node. We further assume that there are vulnerabilities that can be exploited by the attacker to give them the root privilege. We use the information from the reported vulnerabilities and rankings, which are populated using the vulnerability databases, such as the National Vulnerability Database (NVD)~\cite{mell2006common}. Our additional assumptions are as follows:
(1) The cloud provider permits virtual machine-live migration (VM-LM) for cloud-band nodes; (2) the VM-LM downtime is negligible; (3) the cloud provider purchases enough licenses for a back-up OS; and (4) the VMs' OS can be replaced with other variants if needed.

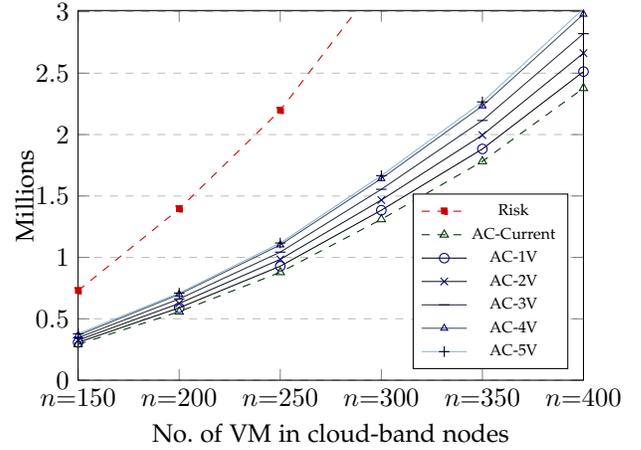
\begin{figure}[t]
	\centering
		\begin{tikzpicture}
		\begin{axis}[
	height=6.5cm,
	width=8.3cm,
		y label style={at={(axis description cs:-0.07,.5)},anchor=south},
		legend style={font=\scriptsize ,nodes={scale=0.85},mark options={scale=1}}, 
	xlabel={No. of VM in cloud-band nodes},
	ylabel={Millions},
	xmin=1, xmax=6,
	ymin=0, ymax=3.01,
	xtick={0,1,2,3,4,5,6},
	ytick={0,0.5,1,1.5,2,2.5,3},
	xticklabels={,$n$$=$$150$,$n$$=$$200$,$n$$=$$250$,$n$$=$$300$,$n$$=$$350$,$n$$=$$400$},
		legend pos=south east,
		ymajorgrids=true,
		grid style=dashed,
		]
		
	\addplot[
	color=red!80!black,
	dashed,mark=square*,mark options={scale=0.6}]
	table [x=VM, y=Risk, col sep=comma]{data-charts.csv};
	
	\addplot[
	color=green!20!black,
	dashed,mark=triangle,mark options={scale=0.9,solid}]
	table [x=VM, y=AC-C, col sep=comma]{data-charts.csv};	
	
	\addplot[
	color=babyblueeyes!5!black,
	mark=o,mark options={scale=0.9,solid,color=blue!50!black}]
	table [x=VM, y=AC-1, col sep=comma]{data-charts.csv};	
		
	\addplot[
	color=babyblueeyes!15!black,
	mark=x,mark options={scale=1,solid,color=blue!40!black}]
	table [x=VM, y=AC-2, col sep=comma]{data-charts.csv};
	
		\addplot[
	color=babyblueeyes!25!black,
	mark=-,mark options={scale=1,solid,color=blue!30!black}]
	table [x=VM, y=AC-3, col sep=comma]{data-charts.csv};
	
		\addplot[
	color=babyblueeyes!40!black,
	mark=triangle,mark options={scale=0.75,solid,color=blue!50!black}]
	table [x=VM, y=AC-4, col sep=comma]{data-charts.csv};
	
		\addplot[
	color=babyblueeyes!90!black,
	mark=+,mark options={scale=1,solid,color=blue!10!black}]
	table [x=VM, y=AC-5, col sep=comma]{data-charts.csv};
		
		
		\legend{$\text{Risk}$\\
			$\text{AC-Current}$\\
			$\text{AC-1V}$\\
			$\text{AC-2V}$\\
			$\text{AC-3V}$\\
			$\text{AC-4V}$\\
			$\text{AC-5V}$\\}				
		
		\end{axis}
		\end{tikzpicture}
	\caption{Comparison of $AC$ and $Risk$ values obtained after deploying the {Diversity} technique on the multiple VMs with the highest {betweenness} values for the cloud-band example with various node sizes.}
	\label{fig:m-diversity}
\end{figure}

The HARM can be used to compute the security metrics, such as $Risk$, $AC$, and $RoA$. 
To provide a comprehensive security overview, the computation will incorporate the analysis of all possible attack paths that are calculated using the upper layer of the HARM. For the reliability analysis, we use the SHARPE software package~\cite{sahner2012performance}. We assume that the attack rate for the cloud-band models follows an exponential function with an average value of $0.2$ (i.e., one attack per every five hours). In addition, we compute the $Reliability$ values during a $10$-hour period. We compute the defined security metrics of the current cloud-band system before deploying the MTD techniques. We compute these metrics for different cloud-band sizes ranging from 150 VMs up to 400 VMs in the cloud.
{Figure~{\ref{fig:current-metrics}} shows the current security posture of the cloud based on different nodes, it shows that when the size of the cloud increases, the three metrics increase too, but the increase of the metrics is in a linear fashion.}
In the following sections, we compare the results obtained from deploying the MTD techniques on the cloud-band against the current security posture of the cloud to investigate the effectiveness of the MTD techniques.

\subsection{Diversity on Multiple VMs}\label{diversity}

\begin{figure*}[h]
	\begin{subfigure}{0.33\textwidth}
		\includegraphics[height=4cm, width=6cm]{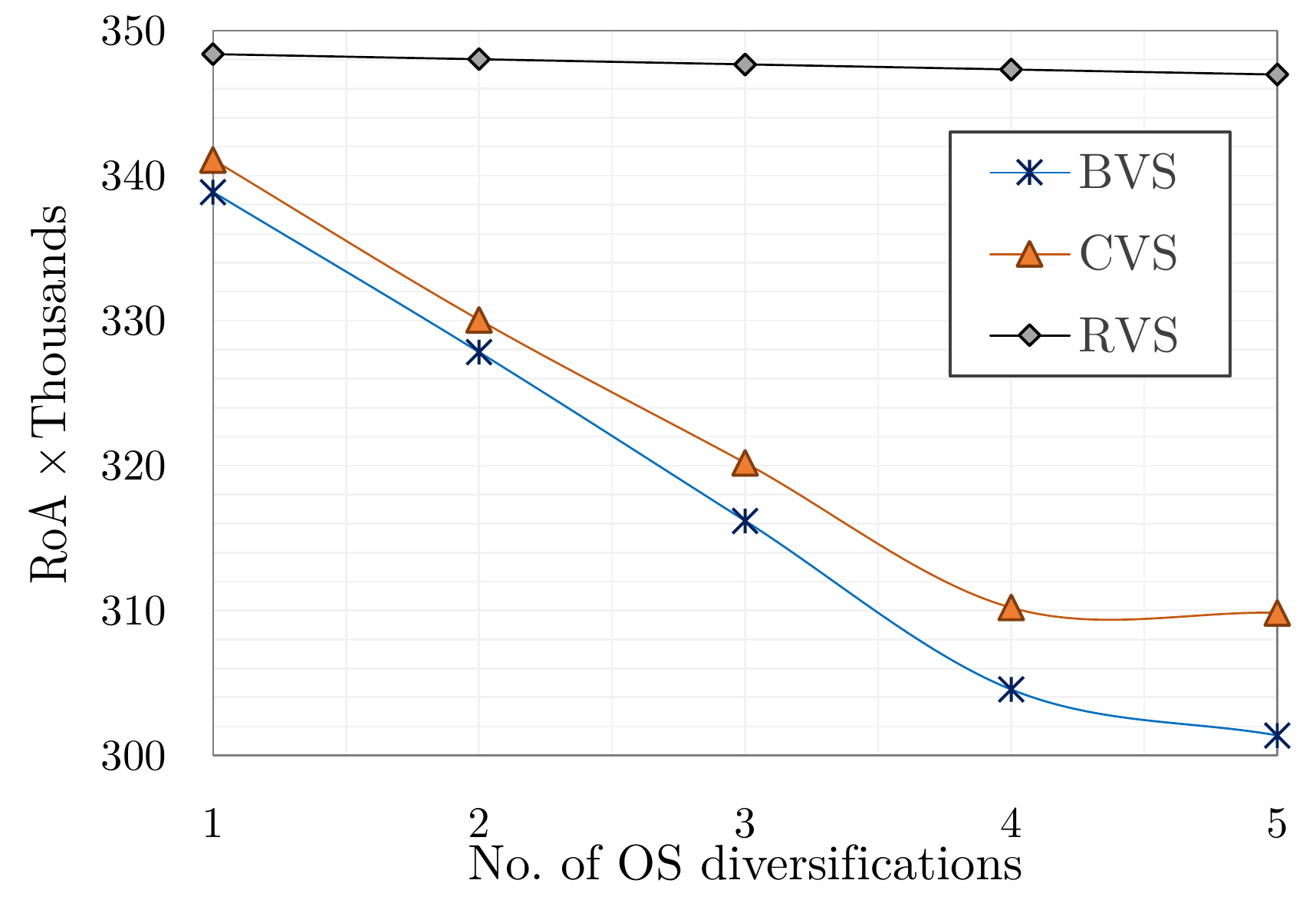}
		\caption{$n=200$}
		\label{DR200}
	\end{subfigure}
	\begin{subfigure}{0.33\textwidth}
		\includegraphics[height=4cm, width=6cm]{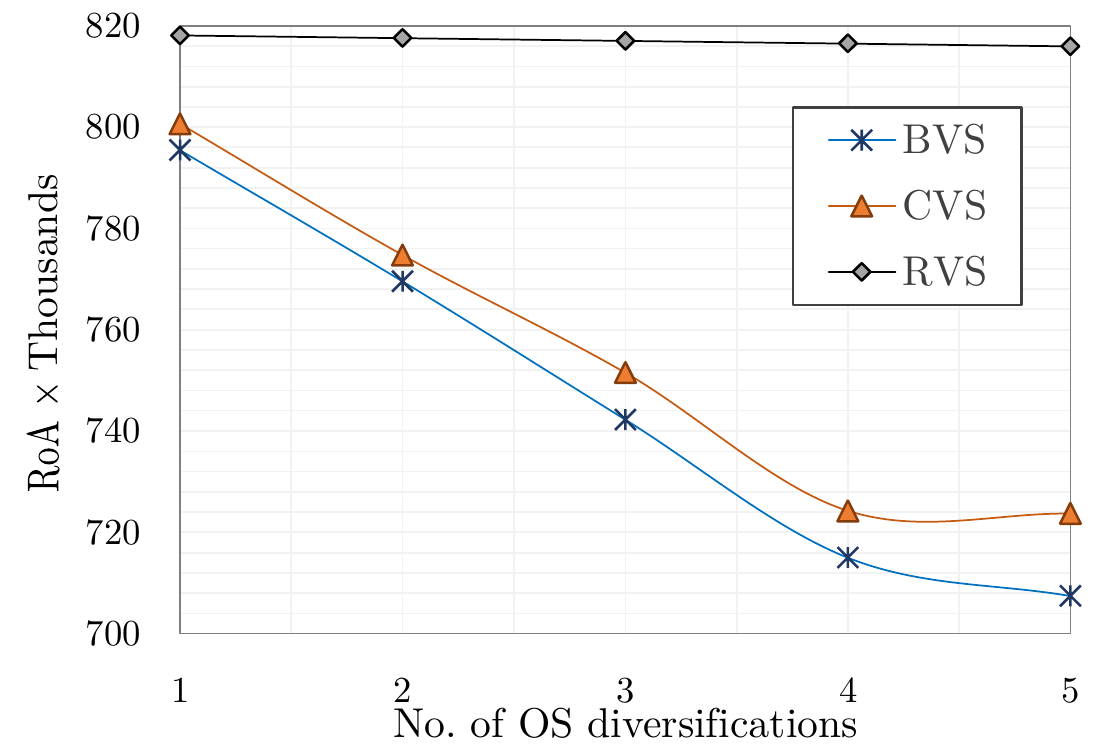}
		\caption{$n=300$}
		\label{DR300}
	\end{subfigure} 
	\begin{subfigure}{0.33\textwidth}
		\includegraphics[height=4cm, width=6cm]{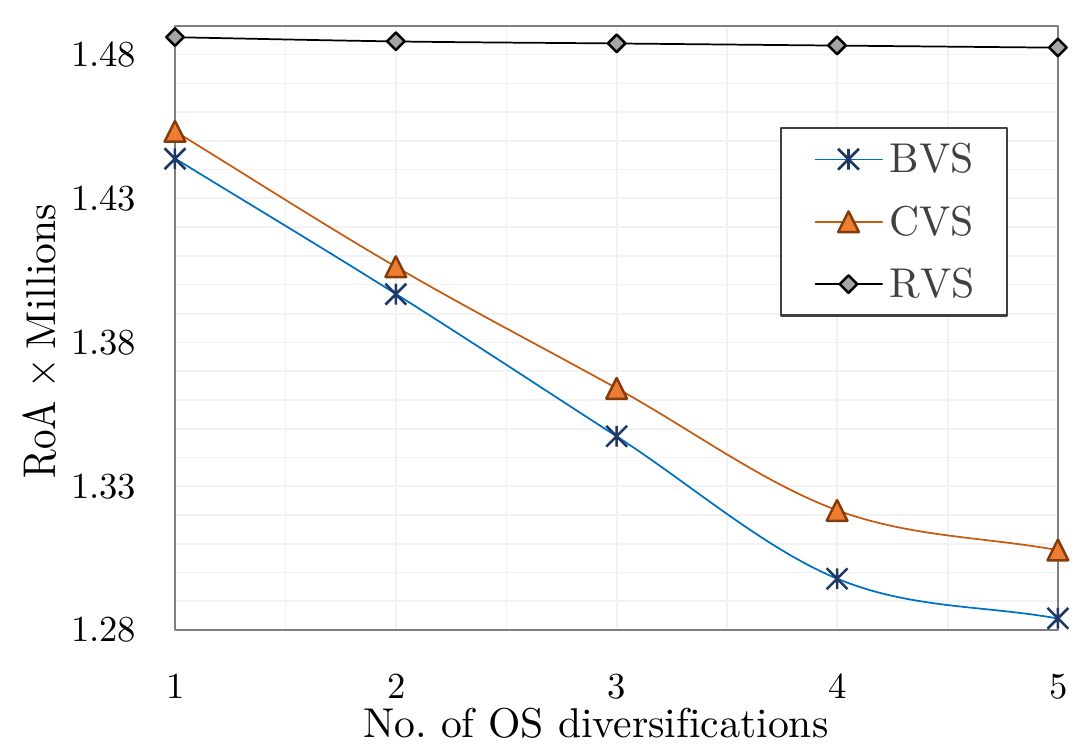}
		\caption{$n=400$}
		\label{DR400}
	\end{subfigure}
	\caption{Comparing the result of $RoA$ metrics after deploying the {Diversity} technique on multiple nodes selected based on three different criteria on the cloud-band with various numbers of VMs.}
	\label{fig:D-RoA}
\end{figure*}

The Diversity MTD techniques play a crucial role in forcing attackers to increase their efforts. Attackers must spend time and money to gain sufficient knowledge to discover and exploit the vulnerabilities of a system (i.e., the network's component, VM, service, and OS). Any sudden changes in those components can confuse the attacker, which may, in turn, increase the time and effort needed to carry out the attack. We consider using the OS diversification method as the main technique for deploying the Diversity technique. In this section, we consider deploying Diversity techniques on multiple VMs (i.e., for different subsets of VMs). To evaluate the effectiveness of the Diversity MTD technique, we deploy OS diversification on multiple VMs based on three selection criteria: (i) betweenness VM selection (\emph{BVS}), (ii) closeness VM selection (\emph{CVS}), and (iii) random VM selection (\emph{RVS}). The \emph{RVS} method selects a set of random VMs in the cloud. The \emph{BVS} method selects the set of VMs based on their higher betweenness ranks, which is one of the NCM measures. Similarly, the \emph{CVS} method uses the closeness ranks of the VMs in the cloud-band.

Deploying the Diversity MTD technique preserves the upper layer of the HARM. Here, we assume that the cloud provider has up to five OS variants as back-up. Note that increasing the OS variants raises the cloud provider's costs (i.e., the costs of purchasing OS licenses for the OS variants). For simplicity, we do not consider the cloud provider's costs in this section.

\begin{figure}[t]
	\begin{subfigure}{0.50\textwidth}
		\centering
		\begin{tikzpicture}
		\begin{axis}[
		ybar,
		height=6.5cm,
		width=8.4cm,
		bar width=5.5pt,
		y label style={at={(axis description cs:-0.07,.5)},anchor=south},
		font=\small,
		xlabel={No. of OS Diversification},
		ylabel={Millions},
		xmin=0.5, xmax=5.5,
		ymin=0, ymax=3.22,
		xtick={0,1,2,3,4,5},
		ytick={0,0.5,1,1.5,2,2.5,3},
		xticklabels={,$1V$,$2V$,$3V$,$4V$,$5V$,},
		yticklabels={$0$,$0.5$,$1$,$1.5$,$2$,$2.5$,$3$,},
		legend pos=north west,
		ymajorgrids=true,
		yminorgrids=true,
		grid style=dashed,
		]
		
		\addplot[
		color=blue!35!black,fill=blue!20!white,
		]
		table [x=VM, y=AC-all, col sep=comma]{data2.csv};

		\addplot[
		color=babyblueeyes!80!black,fill=babyblueeyes!40!white,
		]
		table [x=VM, y=RoA-all, col sep=comma]{data2.csv};	
		
		\addplot[
		color=red!80!black,fill=red!20!white,
		]
		table [x=VM, y=AC-D, col sep=comma]{data2.csv};	
		
		\addplot[
		color=bananamania!50!black,fill=bananamania,
		]
		table [x=VM, y=RoA-D, col sep=comma]{data2.csv};
		
		\legend{\footnotesize{S+R+D, AC}\\
			\footnotesize{S+R+D, RoA}\\
			\footnotesize{D-Only, AC}\\
			\footnotesize{D-Only, RoA}\\},
		
		\end{axis}
		\end{tikzpicture}
		\caption{}
		\label{fig:comb-Metrics}
	\end{subfigure}
	
	\begin{subfigure}{0.5\textwidth}
		\centering
		\begin{tikzpicture}
		\begin{axis}[
	height=6.5cm,
	width=8.4cm,
		y label style={at={(axis description cs:-0.08,.5)},anchor=south},
		legend style={font=\footnotesize ,nodes={scale=0.85},mark options={scale=1}}, 
	xlabel={Time (Hour)},
	ylabel={Millions},
	xmin=0, xmax=10,
	ymin=0, ymax=1,
		xtick={0,1,2,3,4,5,6,7,8,9,10},
		ytick={0,0.1,0.2,0.3,0.4,0.5,0.6,0.7,0.8,0.9,1},
		legend pos=south west,
		ymajorgrids=true,
		xmajorgrids=true,
		grid style=dashed,
		]
		
	\addplot[
	color=red!80!black,
	dashed,mark=square,mark options={scale=0.9,solid}]
	table [x=T, y=RC-150, col sep=comma]{data-R.csv};
	
	\addplot[
	color=green!20!black,
	mark=+,mark options={scale=0.9,solid,color=black}]
	table [x=T, y=RC-400, col sep=comma]{data-R.csv};	
	
	\addplot[
	dashed,color=blue!80!black,
	mark=o,mark options={scale=1.1,solid,color=blue!40!black}]
	table [x=T, y=RM-150, col sep=comma]{data-R.csv};	
		
	\addplot[
	color=green!20!black,
	mark=-,mark options={scale=1.2,solid,thick}]
	table [x=T, y=RM-400, col sep=comma]{data-R.csv};

		\legend{Current $n$=$150$\\
			Current $n$=$400$\\
			S+D+R $n$=$400$\\
			S+D+R $n$=$400$\\}				
		\end{axis}
		\end{tikzpicture}
		\caption{}
		\label{fig:comb-A}
	\end{subfigure}
	\caption{(a) Comparing the results of $AC$ and $RoA$ for D-Only and S+D+R based on different number of Diversity techniques from 1V to 5V. (b) Comparing the $Reliability$ values after S+R+D with the $Reliability$ values for the current cloud-band for $n=150,~n=400$.}
	\label{fig:combined}
\end{figure}
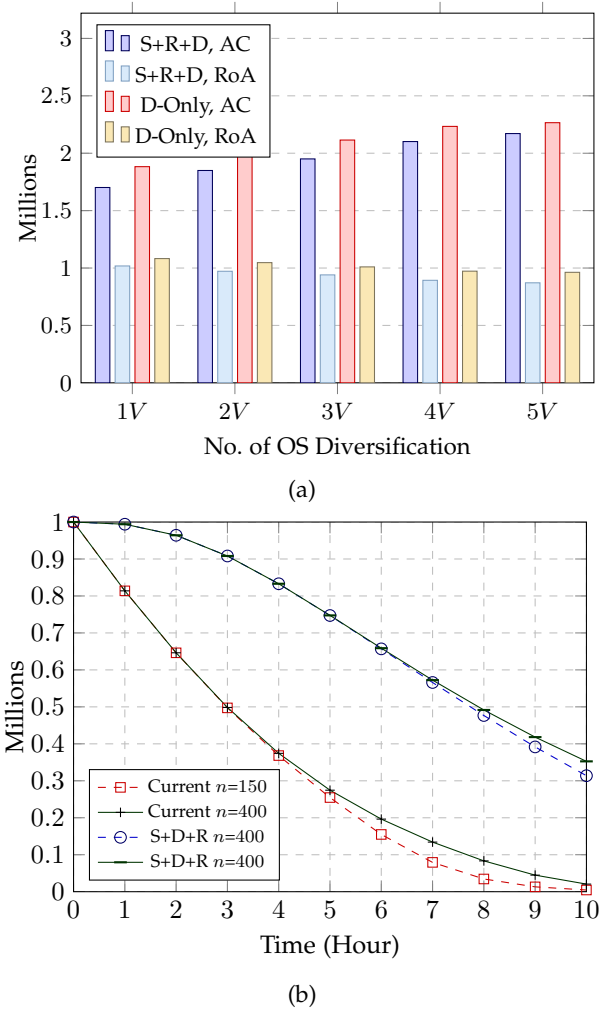

Figure~\ref{fig:m-diversity} shows the $AC$ and $Risk$ metrics after deploying the Diversity technique on multiple VMs ranked according to higher betweenness values. We denote $xV$ as deploying new variants on $x$ selected VMs, respectively. Then, $5V$ represents deploying different OS variants on the five selected VMs. We observe that deploying Diversity on the VMs with higher betweenness ranks results in a larger increase in the costs for the attacker, and raises the $AC$ value. This increase is greater if we add to the number of OS variants and deploy the Diversity technique on multiple nodes. Deploying OS diversification by assigning $5V$ in a cloud band with 400 VMs increases the $AC$ value from around 2.3 million to three million, while the increase for $1V$ does not go beyond 2.5 million.

We do not observe a any changes in the $Risk$ values after deploying the Diversity MTD techniques. {The $Risk$ value remains almost steady after increasing the OS variants, but it still increases sharply after increasing the cloud-band nodes; see Figure~{\ref{fig:m-diversity}}. We could also consider how the metrics normalized by dividing by $n$ change when $n$ increases. Doing so indicates that the attacker has to spend more resources per VM if the defender has greater scope for diversity.}

Figure~\ref{fig:D-RoA} compares the results of the $RoA$ metric by applying the Diversity technique to three VM selection groups in different cloud-band sizes: BVS, CVS, and RVS. These observations indicate that deploying the Diversity technique on a set of VMs selected using BVS provides better results than the other groups. The values of $RoA$ for the RVS groups have very gentle decrements, while the OS diversification increases. However, the $RoA$ values for both the BVS and CVS groups decrease sharply when the OS diversification numbers increase. Other results for different cloud-band nodes show the same trend, but, as was expected, the $RoA$ values for the cloud-band that include more VMs are higher (i.e., $RoA$ values of the cloud-band with 400 VMs are between one and 1.5 million, Figure~\ref{DR400}, while this rate is between 300,000 and 350,000 for the cloud-band including 200 VMs; Figure~\ref{DR200}).

\begin{figure*}[t]
	\centering
	\begin{subfigure}[b]{0.24\textwidth}
		\centering
		\includegraphics[height=1cm]{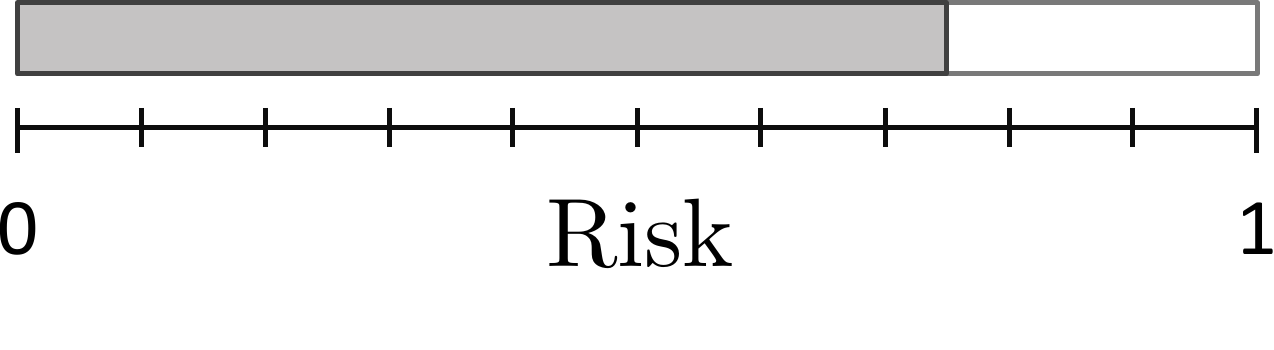}
	\end{subfigure}
	\begin{subfigure}[b]{0.24\textwidth}
		\centering
		\includegraphics[height=1cm]{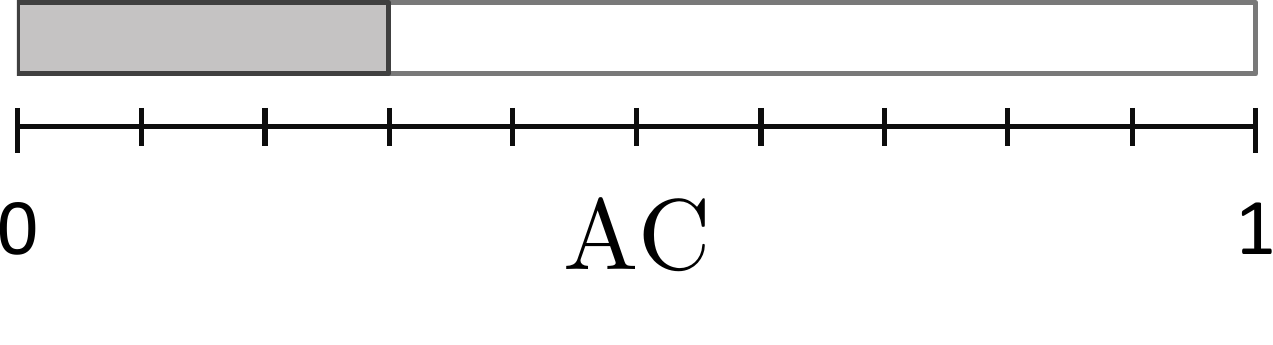}
	\end{subfigure}
	\begin{subfigure}[b]{0.24\textwidth}
		\centering
		\includegraphics[height=1cm]{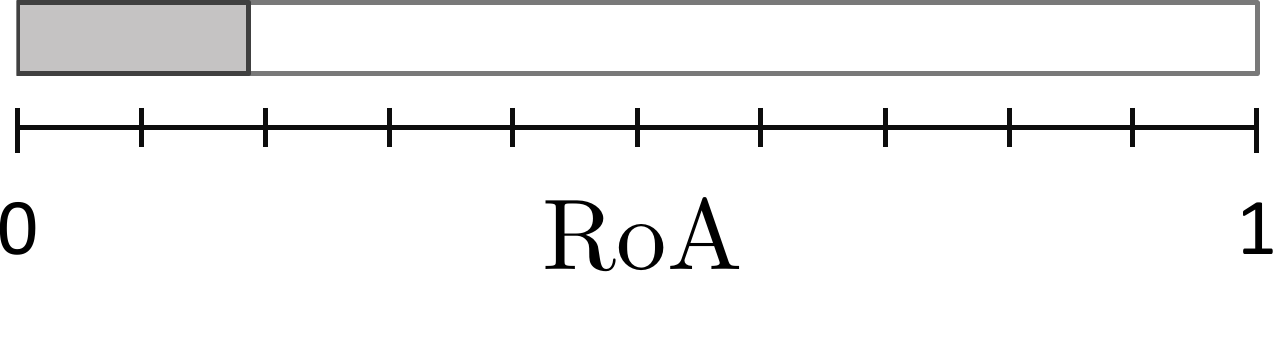}
	\end{subfigure}
	\begin{subfigure}[b]{0.24\textwidth}
		\centering
		\includegraphics[height=1cm]{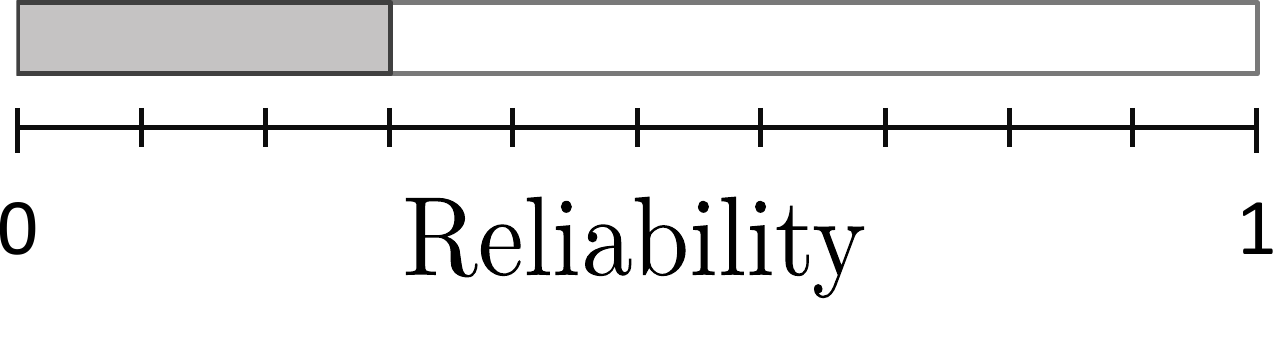}
	\end{subfigure}
	\\
	\begin{subfigure}[b]{0.24\textwidth}
		\centering
		\includegraphics[height=1cm]{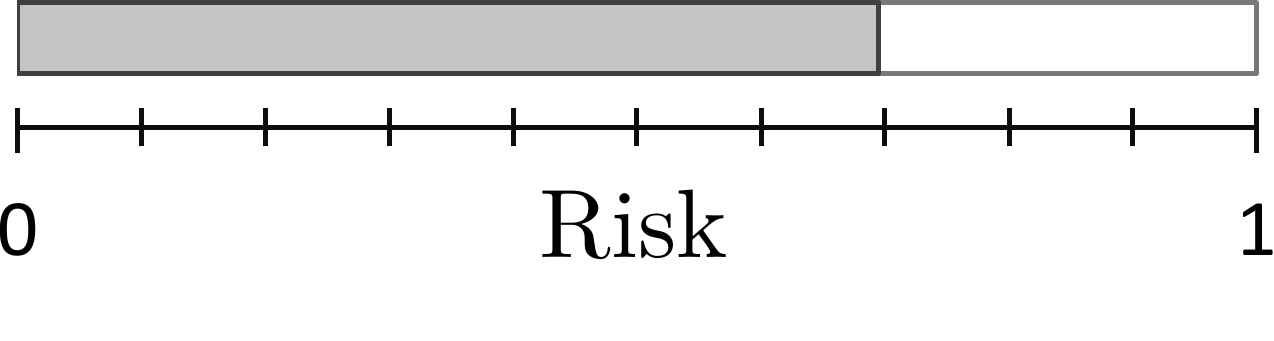}
	\end{subfigure}
	\begin{subfigure}[b]{0.24\textwidth}
		\centering
		\includegraphics[height=1cm]{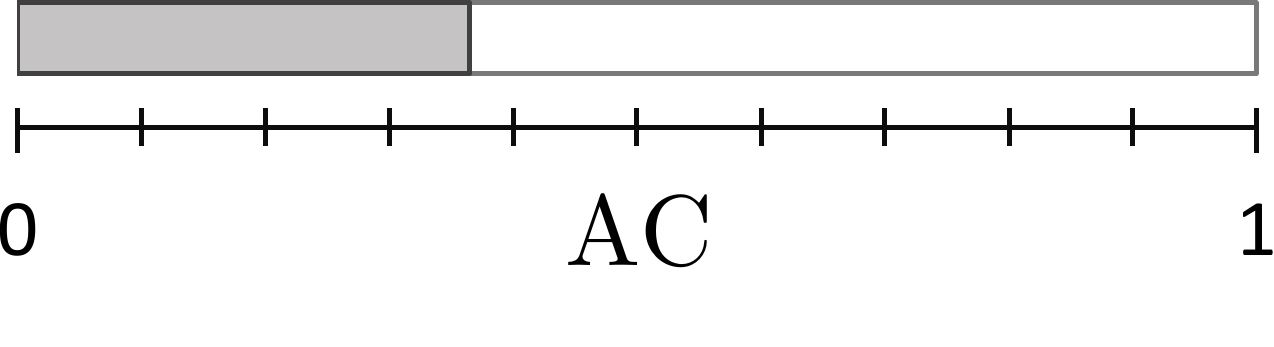}
	\end{subfigure}
	\begin{subfigure}[b]{0.24\textwidth}
		\centering
		\includegraphics[height=1cm]{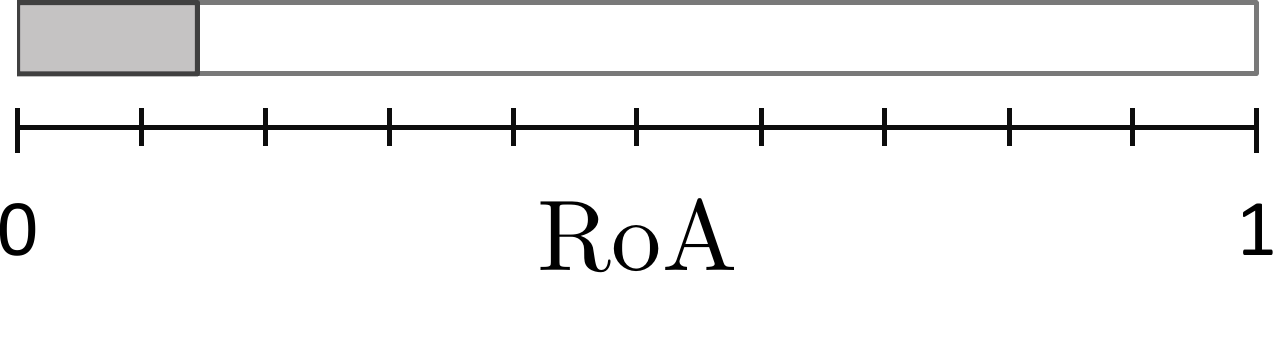}
	\end{subfigure}
	\begin{subfigure}[b]{0.24\textwidth}
		\centering
		\includegraphics[height=1cm]{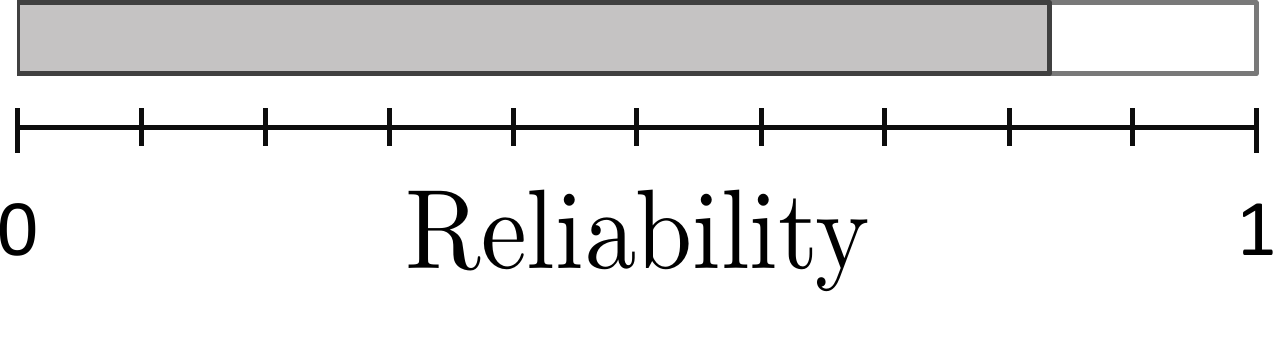}
	\end{subfigure}
	\caption{Line chart comparing normalized metrics of the cloud-band with $n=350$ before and after deploying S+D+R: the upper line charts show the current cloud-band and the lower line charts show the metrics after deploying MTD techniques.}
	\label{fig:lines}
\end{figure*}

\subsection{Combining Shuffle, Diversity, and Redundancy}\label{S+D+R}
We combine the main three MTD techniques, Shuffle, Diversity, and Redundancy (S+D+R), to investigate whether this approach enhances the security of the cloud. It is important to quantify the effects of the combined MTD techniques, and to compare them with the current security level of the system. In order to combine the three techniques, we set combination criteria based on the results obtained from previous sections.

The results discussed in Section~\ref{diversity} showed that OS diversification on the VMs having higher {betweenness} (grouped in \emph{BVS}) has a better effect on security metrics. They also revealed that increasing the numbers of OS variants increases $AC$ and decreases $RoA$. Based on those results, we only consider the \emph{BVS} group for deploying the {Diversity} technique in this section. Moreover, based on the results of previous studies reported in~\cite{alavizadeh2021evaluating} for combining the {Shuffle} and {Redundancy} techniques, we deploy the {Shuffle} technique on the most suitable VMs, which can be found by analyzing only the top 10\% of the VMs with higher values of {betweenness}. Finally, we deploy the {Redundancy} on the DB or on VMs connected to the DB (target) to increase the reliability of the cloud and availability of the DB against DDoS attacks. 

Figure~\ref{fig:comb-Metrics} compares the $AC$ and $RoA$ metrics after deploying S+R+D with $5V$ with those of the Diversity technique (D-Only). We observe that the values of $AC$ in D-Only are lower than the values of $AC$ in S+D+R. However, the corresponding $RoA$ values for S+D+R are also lower in D-Only, which shows that the attacker is less likely to attack again. Figure~\ref{fig:comb-A} compares the $Reliability$ values of the cloud-band with 150 and 400 VMs before and after deploying the S+D+R technique. We select the boundaries of $n=150$ and $n=200$ as the other values of $n$ fall between these two boundaries. We can observe how the $Reliability$ values increase after deploying S+D+R. After passing 10 hours of an assumed attack with $\alpha$$=$$0.2$, the cloud-bands that are secured with S+D+R are around 40\% reliable, while this rate reaches almost 0\% for the current cloud-bands on which no MTD techniques are deployed.

We deployed the S+D+R technique with $5V$ on the cloud-band and evaluate the results. {Figure~{\ref{fig:lines}} provides an overview of the changes in the security metrics by comparing the normalized values for the security metrics (between zero and one) as the results of deploying S+D+R on the cloud-band with 350 VMs.} It is clear that the security metrics are improved after deployment.

Nevertheless, the combination of MTD techniques can be quantified based on the security levels expected by the cloud providers of network administrators. As deploying MTD techniques can be costly, it is important for cloud providers to understand the trade-offs between security and economic considerations.

\section{Economic Metrics for MTD techniques}\label{Economic}
\begin{table*}[ht]
	\centering
	\caption{VM assets and vulnerabilities (Note that $vm_{10}$ is the target VM and includes PHI records.)}
	\label{Vuls}
	\begin{tabular}{@{}lllllllll@{}}
		\toprule
		\multirow{2}{*}{VMs}           & \multirow{2}{*}{OS ($\theta$)} &  \multirow{2}{*}{\begin{tabular}[c]{@{}l@{}}Asset Value\\ (AV) ($\$$)  \end{tabular}} & \multicolumn{6}{l}{Vulnerabilities (V)}                                                                                \\ \cmidrule(l){4-9} 
		&                                &                                               &  V-ID        & CVE-ID   &Threat   & Exploitability& AC & EF (\%) \\ \midrule
		\multirow{3}{*}{$vm_1$$-$$vm_5$} & \multirow{3}{*}{\textbf{W}in10}         & \multirow{3}{*}{500}                                                             & $\nu_{1,W}$ & CVE-2018-8490 & Remote                                                         & 0.17   & 1.6 & 0.6    \\
		&                                                                                                          &                                                                                  & $\nu_{2,W}$ & CVE-2018-8484 & Privilege Escalation                                                       & 0.18 & 2.2 & 0.59    \\
		&                                                                                                  &                                                                                  & $\nu_{3,W}$ & CVE-2018-0784
		& Privilege Elevation                                                       & 0.28& 1.2 & 0.59    \\ \cmidrule(l){4-9} 
		\multirow{3}{*}{$vm_6$$-$$vm_9$} & \multirow{3}{*}{\textbf{L}inux}          & \multirow{3}{*}{480}                                                             & $\nu_{1,L}$ & CVE-2018-14678 & DDoS                                                       & 0.18  & 2.2 & 0.59   \\
		&                                                                                                         &                                                                                  & $\nu_{2,L}$ & CVE-2018-14633 & DDoS \& Remote                                                     & 0.22  & 3   & 0.47      \\
		&                                                                                                         &                                                                                  & $\nu_{3,L}$ & CVE-2017-15126 & Use After Free (UAF)                                                         & 0.22  & 1.9 & 0.59      \\ \cmidrule(l){4-9} 
		\multirow{2}{*}{$vm_{10}$}     & \multirow{2}{*}{\textbf{L}inux}                                                             & \multirow{2}{*}{10000}                                                           & $\nu_{1,L}$ & CVE-2018-14678 & DDoS                                                         & 0.18  & 2.2 & 0.59   \\
		&                                                                                                          &                                                                                  & $\nu_{2,L}$ & CVE-2018-14633 & DDoS \& Remote                                                      & 0.22 & 3   & 0.47     \\ \cmidrule(l){4-9} 
		Back-up OS                      & \textbf{F}edora                                     & 450                                                                              & $\nu_{1,F}$ & CVE-2014-1859 & Symlink attack & 0.18 & 4.5 & 0.3   \\ \bottomrule
	\end{tabular}
\end{table*}

Although security metrics reveal different dimensions of a cloud's security posture, investigating the economic aspects of the deployment of MTD techniques is also crucial. {Many studies have weighed the costs and benefits of investments in security using different models, such as } \cite{abdallah2020behavioral, hota2018game, gordon2002economics}. {In }\cite{abdallah2020behavioral}{, the authors utilized an attack graph that analyzed the impacts of behavioral decision-making on security against the costs of security investments using a game-theoretic model. By contrast, in }\cite{hota2018game}{, the authors conducted a comprehensive evaluation of the benefits of security investments using a multi-defender game-theoretic model. In particular, they considered using a MTD technique to make the defensive model dynamic by applying changes in the configuration that fell into the Shuffle MTD category. An analysis that used a similar approach to evaluating multiple MTD techniques was presented in~}\cite{connell2017framework}{. In this study, the authors utilized three MTD techniques -- namely, service rotation, IP rotation, and address space layout randomization (ASLR) techniques -- and quantified the benefits of using multiple MTDs versus using a single MTD. However, this evaluation was based on Shuffle techniques only, as all rotation and randomization MTD techniques are categorized as Shuffle techniques~}\cite{cho2020toward}{. Thus, evaluations of MTD techniques in the cloud that are based on both security and economic metrics, and evaluations of the Diversity technique that seek to find an optimal diversity assignment solution, are still missing in the current literature.}

Here, we compute various economic metrics to provide different perspectives on MTD deployment scenarios based on a cloud example.

\subsection{A Case Study Based on the E-Health Cloud Model}
We consider a case study based on the personal health information (PHI) of patients, including their medical histories, located in a private personal health cloud (PHC), as shown in Figure~\ref{fig:Cloud-Model}. We assume that the attacker can use the vulnerabilities of the cloud's components (e.g., VMs) to get into the cloud and to find a path to the PHI database. In this section, we aim to evaluate the effectiveness of MTD techniques in terms of their economic metrics. We utilize the HARM generated for the e-health cloud model as presented in Figure~\ref{fig:HARM1}. Table~\ref{Vuls} displays the existing vulnerabilities on each VM~\cite{NVD}, together with the asset value (AV) and the exploitability factor (EF), which are required for an economic evaluation. Moreover, we assume that the cloud provider has one back-up OS that can be used for the deployment of the Diversity technique. We further assume that the PHI records are stored in a DB connected to $vm_{10}$, and that any successful attack exploiting $vm_{10}$ will cause significant damage (of $\$10,000$) to the organization due to the loss and/or disclosure of patients health information.

\subsection{Single Loss Expectancy}
The single loss expectancy ($SLE$) measures an organization's expected loss from a single threat~\cite{krutz2001cissp}. $SLE$ can be determined for a cloud based on the asset values (AVs) for each VM, including the costs of maintenance, running the OS, services, DB record values, and applications. The estimated AV for each OS is shown in Table~\ref{Vuls}. We assume that $SLE$ can be calculated for both the VM and the network (cloud) levels. The value of $SLE$ for a VM can be obtained by multiplying the asset value and the maximum percentage of loss for that asset caused by a threat, which is called an exposure factor (EF); see Equation~\ref{eq:sle}.

\begin{equation}\label{eq:sle}
SLE_{vm_i}={\Bigg(1-\prod_{v_{j,\theta} \in V_{i,\theta}}{\Big(1-EF_{v_{j,\theta}}\Big)\Bigg)\times AV_{vm_i}}}
\end{equation}
In Equation~\ref{eq:sle}, the AV consists of the costs associated with running an active VM (i.e., purchasing a license for an OS, applications, DB values, etc.).

The $SLE$ for the cloud ($SLE_c$) including all assets (here, VMs) can be calculated based on Equation~\ref{eq:all-sle}.

\begin{equation}\label{eq:all-sle}
SLE_c = \sum_{ap \in AP} \bigg( \sum_{vm_i \in ap} SLE_{vm_i} \bigg)
\end{equation}

{Note that $AP$ denotes the set of all possible attack paths from an attacker to the DB in the cloud model.}

\subsection{Annual Loss Expectancy}
Annual loss expectancy ($ALE$) can be defined as the expected financial loss due to an attack event, and can be computed by the product of $SLE$ and the annualized rate of occurrence ($ARO$), which represents the estimated number of occurrences of a threat event per year~\cite{krutz2001cissp}. 
\begin{equation}\label{eq:all-PLE}
ALE_c = \sum_{ap \in AP} \bigg( \sum_{vm_i \in ap} SLE_{vm_i} \times ARO_{vm_i} \bigg)
\end{equation}

\subsection{Benefits of Security} 
Many defensive strategies can be adapted to the cloud to either avoid or mitigate the exploitation of or damage to the cloud. In this paper, we evaluate the benefits associated with deploying MTD techniques. Benefits of security ($BS$)~\cite{bistarelli2012evaluation} can be used to show the effects of deploying a single or a combination of defensive techniques. The benefits of security for a cloud $BS_c$ can be computed based on Equation~\ref{eq:BS}. In Equation~\ref{eq:BS}, $ALE_c^{\mu}$ denotes the $ALE$ value of the cloud after deploying MTD techniques. $\mu \subseteq \{S,D,R\}$ denotes a set of MTD used as defensive techniques.
\begin{equation}\label{eq:BS}
BS_c^{\mu} = ALE_c - ALE_c^{\mu} 
\end{equation}	

{Another evaluation measurement that uses the $ALE$ values of the cloud before and after deploying the MTD techniques is the \textit{mitigation factor}. The mitigation factor, denoted by $MF^{\mu}$, shows the ability of the defensive MTD techniques to impair the attack. $MF^{\mu}$ takes values within the range $[0,1]$ as in Equation~{\ref{eq:MF}}. Note that, a larger value of $MF^{\mu}$ is more desirable.}
\begin{equation}\label{eq:MF}
MF^{\mu} = 
\begin{cases}
1-\frac{ALE_c^{\mu}}{ALE_c}, &  \text{if } ALE_c^{\mu} <  ALE_c\\
0, & \text{otherwise}
\end{cases}
\end{equation}

\subsection{Costs of Security}
The costs of security ($CS$) consist of any expenses associated with the use of a defensive security mechanism, such as the costs of deployment, purchases, maintenance, and patching~\cite{bohme2010security}. In addition, the costs of security can also include indirect costs, such as the costs associated with a loss of productivity due to system downtime, or of having to perform other security-related activities, such as training. However, for simplicity, we make the following assumptions regarding the costs of security. We assume that the unit cost of deploying the Shuffle technique for a given VM is $\$20$ per operation, which includes the costs of employing experts and the loss of productivity. We also assume that the unit cost of deploying the Diversity technique on a VM such that a given VM is replaced with the back-up OS (Fedora in Table~\ref{Vuls}) is $\$55$ per operation, which includes the costs of experts, performing maintenance, and the loss of productivity for a given VM for an operation per year~\cite{rabai2013cybersecurity,sonnenreich2006return}.

\subsection{Return on Security Investment}
The overall benefits of the selected defensive MTD strategies relative to the costs of implementation can be evaluated using the return on security investment ($RoSI$). $RoSI$ can be used to evaluate the profitability of a defensive investment relative to the costs, as formulated in Equation ~\ref{eq:RoSI}. 
\begin{equation}\label{eq:RoSI}
RoSI_c^{\mu} = \frac{BS_c^{\mu}-CS^{\mu}}{CS^{\mu}}
\end{equation}

\subsection{Shuffle Evaluation}\label{MTD-Shuffle}
In this section, we evaluate the Shuffle technique based on economic metrics using the PHC cloud example. We propose a VM placement strategy based on the shortest path in the upper layer of the HARM. We aim to enhance the migration scenarios for the Shuffle technique by using the shortest path rather than a random VM placement strategy. The idea behind selecting the shortest path strategy for the VM placement is to take advantage of the shortest attack path (SAP)~\cite{yusuf2017composite} alongside other benefits of the Shuffle technique. 

\vspace{1mm}
\noindent {\textbf{Migration Strategy.}}
We propose the shortest path injection approach, in which a selected VM can be moved and connected to the VMs located in the shortest path in the upper layer of the HARM. Since there may be more than one shortest path, we utilize a migration strategy to find the most critical shortest path, as in Equation~\eqref{eq:T}.

Let $SAP=\{sp_1, sp_2, sp_3, \dots, sp_q\}$ be a set of possible shortest paths existing in the upper layer of the HARM. Then, we define the strategy $T_{sp}$ as a selected shortest path with lower in-degree values, as follows.

\begin{equation}\label{eq:T}
T_{sp}=\min_{sp \in SAP}\Big(\sum_{vm_i \in sp} \sum_{j \leq n} a_{ij}\Big)
\end{equation}

We deploy the Shuffle techniques on all VMs in the upper layer of the HARM, and evaluate the effectiveness of each migration scenario. We examine both the security and the economic metrics in order to find the best deployment scenario. {Note that because we lack real data to estimate $ARO$, we assume in all our evaluations that the $ARO$ value is one  (similar to~}\cite{bistarelli2012evaluation}{).} Table~\ref{TS} displays the results of deploying the Shuffle technique on the VMs based on the $Risk_c$, $AC_c$, and $RoA_c$ security metrics; and on the $ALE_c$, $BS_c$, and $RoSI_c$ economic metrics. Table~\ref{TS} shows that the most promising approach based on the security metrics is deploying the Shuffle technique on VM $vm_5$, which yields the lowest $Risk_c$ and $RoA_c$ values. The table also shows that that the most promising approach based on the economic metrics is deploying the Shuffle technique on VM $vm_9$, which yields higher $RoSI_c$ values ($RoSI_c$$=$$2631$) than the other deployment scenarios. However, when the VM $vm_9$ is selected for the Shuffle technique, it still yields adequate results for $RoA$ of about $87$ (the second highest $RoA_c$ values). Thus, it appears that the cloud provider can incorporate both $RoA$ and $RoSI$ into the decision-making process, and prioritize them based on their values.

\begin{table}[t]
	\centering
	\scriptsize
	\caption{The results of deploying the Shuffle technique on each VM in the cloud}
	\label{TS}
	\resizebox{0.49\textwidth}{!}{%
	\begin{tabular}{@{}p{0.5cm}lllllll@{}}
		\toprule
		\multirow{2}{*}{VM ID} & \multicolumn{3}{l}{Security Metrics}      & \multicolumn{3}{l}{Economic Metrics} \\ \cmidrule(l){2-8} 
		& $Risk_c^S$ & $AC_c^S$ & $RoA_c^S$                     & $ALE_c^S ($\$$)$        & $BS_c^S$ ($\$$) & $RoSI_c^S$ & $MF^S_{(\%)}$     \\ \midrule
		\begin{filecontents*}{data1.csv}
order,VM,RD,ACD,RoAD,SLED,ALED,BSD,RoSID,MFD,MFDA,RS,ACS,RoAS,SLES,ALES,BSS,RoSIS,MFS,MFSA
1,$\text{vm}_5$,\textbf{167.31},\textbf{253},\textbf{105.19},157266,\textbf{157266},\textbf{2928},\textbf{52.24},0.982,1.8,\textbf{118.35},136.7,\textbf{77.51},113878,113878,46315,2315,0.711,28.9
2,$\text{vm}_6$,174.27,236.6,117.37,157867,157867,2327,41.31,0.984,1.6,126.14,134.7,87.19,114913,114913,45281,2263,0.717,28.3
3,$\text{vm}_9$,173.62,239.2,116.83,157701,157701,2493,44.33,0.983,1.7,117.65,124.7,81.7,\textbf{107563},\textbf{107563},\textbf{52630},\textbf{2631},0.671,32.9
4,$\text{vm}_4$,171.32,239.8,110.12,157998,157998,2196,38.93,0.986,1.4,128.97,142.5,87.15,115712,115712,44482,2223,0.722,27.8
5,$\text{vm}_2$,173.33,233.2,112.59,158364,158364,1830,32.27,0.989,1.1,135.58,147.3,92.66,116912,116912,43282,2163,0.73,27.1
6,$\text{vm}_3$,177.35,220,117.52,159096,159096,1098,18.96,0.993,0.7,149.98,163.5,102.27,131045,131045,29149,1456,0.818,18.2
7,$\text{vm}_1$,171.32,239.8,110.12,157998,157998,2196,38.93,0.986,1.4,140.54,150.9,96.79,117812,117812,42382,2118,0.735,26.5
8,$\text{vm}_7$,175.57,231.4,118.44,158199,158199,1994,35.26,0.987,1.3,145.73,158.5,99.52,130178,130178,30015,1500,0.812,18.8
9,$\text{vm}_8$,178.82,218.4,121.14,159030,159030,1163,20.15,0.992,0.8,182.07,198.3,124.23,159910,159910,283,13,0.998,0.2
\end{filecontents*}
\csvreader[head to column names]{data1.csv}{}
		{\VM & \RS & \ACS& \multicolumn{1}{l|}{\RoAS}                   & \ALES         & \BSS  & \RoSIS & \MFSA             \\ }
		Best&$\text{vm}_5$&$\text{vm}_8$&$\text{vm}_5$&$\text{vm}_9$&$\text{vm}_9$&$\text{vm}_9$&$\text{vm}_9$\\  \bottomrule
	\end{tabular}
	}
\end{table}	

\begin{table}[b]
	\centering
	\scriptsize
	\caption{The results of deploying the Diversity technique on each VM in the cloud}
	\label{TD}
	\resizebox{0.49\textwidth}{!}{
	\begin{tabular}{@{}p{0.5cm}lllllll@{}}
		\toprule
		\multirow{2}{*}{VM ID} & \multicolumn{3}{l}{Security Metrics}      & \multicolumn{3}{l}{Economic Metrics} \\ \cmidrule(l){2-8} 
		& $Risk_c^D$ & $AC_c^D$ & $RoA_c^D$                    & $ALE_c^D ($\$$)$        & $BS_c^D ($\$$)$  & $RoSI_c^D$ & $MF^D_{(\%)}$     \\ \midrule
		\csvreader[head to column names]{data1.csv}{}
		{\VM & \RD & \ACD& \multicolumn{1}{l|}{\RoAD}                & \ALED         & \BSD  & \RoSID & \MFDA             \\ }
		Best&$\text{vm}_5$&$\text{vm}_5$&$\text{vm}_5$&$\text{vm}_5$&$\text{vm}_5$&$\text{vm}_5$&$\text{vm}_5$\\ \bottomrule
	\end{tabular}
	}
\end{table}	
\subsection{Diversity Evaluation}\label{MTD-Diversity}
In Subsection~\ref{diversity}, we evaluated the Diversity technique using security metrics in a large cloud-band model. Here, we extend our evaluation of the Diversity technique by considering economic metrics and optimization. We consider deploying the Diversity technique based on two scenarios: (i) deploying the Diversity technique on single or multiple VMs using a single back-up OS; and (ii) deploying the Diversity technique on multiple VMs using multiple back-up OSs (using an optimization model).

\vspace{1mm}
\noindent {\textbf{Scenario 1.}}
We deploy the Diversity technique only on a single VM through exhaustive search (ES). We evaluate the effectiveness of the Diversity technique by computing the security and economic metrics. Table~\ref{TD} shows the results of deploying the Diversity technique on each VM. The experimental results show that for the security metrics, deploying the Diversity technique on VM $vm_5$ yields the best results in terms of $Risk_c$, $AC_c$, and $RoA_c$. For the economic metrics, it yields the lowest $ALE_c$ , which is $157266$. It also leads to the highest $BS_c$ and $RoSI_c$ values, which are $2928$ and $52.24$, respectively.

\begin{figure}[t]
\begin{tikzpicture}

\begin{axis}[
axis y line*=left,
height=6.5cm,
width=8.3cm,
y label style={at={(axis description cs:-0.1,.5)},anchor=south},
yticklabel style={font=\small},
legend style={nodes={scale=0.85},mark options={scale=1}}, 
xlabel={No. of OS diversification},
ylabel={RoSI values},
xmin=0.5, xmax=9.5,
ymin=20, ymax=80,
xtick={0,1,2,3,4,5,6,7,8,9},
ytick={20,30,40,50,60,70,80},
ymajorgrids=true,
grid style=dashed,
]
\addplot[
black,
mark=square,
]
table [x=VM, y=RoSI-RVS, col sep=comma]{data3.csv};	\label{plot_1_y1}

\addplot[
color=blue,
mark=-,
mark size=2pt,
mark options={ultra thick},
]
table [x=VM, y=RoSI-BVS, col sep=comma]{data3.csv};	\label{plot_1_y2}

\end{axis}

\begin{axis}[
axis y line*=right,
height=6.5cm,
width=8.3cm,
xmin=0.5, xmax=9.5,
axis x line=none,
yticklabel style={font=\small},
legend style={nodes={scale=0.85},mark options={scale=1}}, 
ylabel={CS},
ymin=0, ymax=300,
ytick={0,50,100,150,200,250,300},
legend pos=south east,
]

\addplot[
color=red!80!black,
mark=triangle,
]
table [x=VM, y=CS, col sep=comma]{data3.csv}; \label{plot_1_y3}

\addlegendimage{/pgfplots/refstyle=plot_1_y1}\addlegendentry{CS}
\addlegendimage{/pgfplots/refstyle=plot_1_y2}\addlegendentry{RoSI-RVS}
\addlegendimage{/pgfplots/refstyle=plot_1_y3}\addlegendentry{RoSI-BVS}

\addplot[only marks,blue, mark=star, mark size=4pt]
coordinates{ 
	(4,277)
}; \label{plot_one}
\end{axis}

\end{tikzpicture}
	\caption{Comparing RoSI values obtained after deploying the Diversity technique on various VMs to the CS based on RVS and BVS (the asterisked point shows the optimal solution.)}
\label{fig:Various-D}
\end{figure}
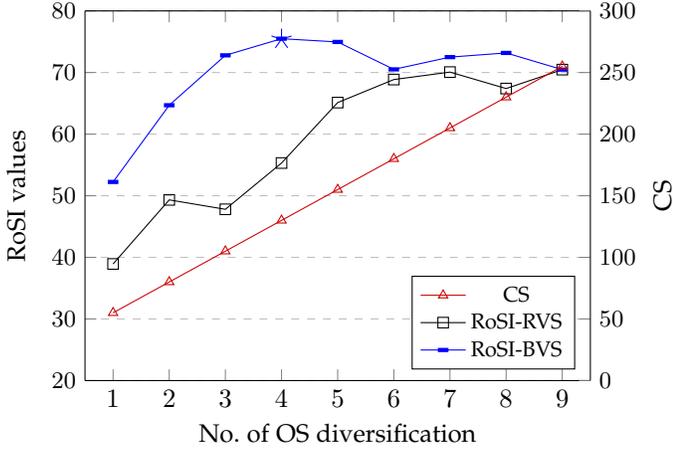

\vspace{1mm}
\noindent {\textbf{Scenario 2.}}
We evaluate the deployment of the Diversity technique on various VMs with a single back-up OS. To do so, we leverage two strategies for finding a set of VMs for deploying the Diversity technique (OS diversification): random VM selection (RVS) and betweenness VM selection (BVS) (as in Subsection~\ref{diversity}). We compare the results of deploying these two strategies on the PHC cloud example while focusing on the return on the security investment and the cost of security. We aim to find a trade-off between the number of OS diversification variants (using the same back-up OS) on a set of VMs and the CS, while maximizing the RoSI. Figure~\ref{fig:Various-D} compares the results of deploying the Diversity technique on various numbers of VMs (from one to nine OS diversification variants) for $RoSI$ values based on RVS and BVS strategies. The results show that deploying the Diversity technique on the four VMs with the highest betweenness values reaches a peak and yields the best $RoSI$ values, while the cost of security remains between $\$100$ and $\$150$. By contrast, the results based on the RVS strategy suggest that deploying nine back-up OSs increases the $RoSI$, but incurs the highest cost of security, of more than $\$250$. Moreover, the highest $RoSI$ metric after deploying the BVS strategy is about $75$, while the highest $RoSI$ metric resulting from the RVS strategy is $70$.

\subsection{Optimal Diversity Assignment}\label{optimal}
In this section, we analyze the deployment of the Diversity technique with multiple OS variants on multiple VMs using an optimization model. We model the decision problem of maximizing the expected net benefit by assigning back-up operating systems to existing virtual machines as a mathematical programming model with binary decision variables. In our model, we consider a graph coloring scheme in which each back-up OS variant is assigned a color (c) in the graph, such that no two adjacent VMs are assigned the same color (back-up OS). The aim of this approach is to increase the difficulty for an attacker, who would encounter a different back-up OS in adjacent VMs through the attack path.

We use $\theta=\{1,2,\dots,k=|\theta|\}$ to represent the set of all potential back-up operating systems from which to choose to implement the Diversity technique on some virtual machines.  

We use the binary decision variable $d_{ic} \ \forall i \in VM, c \in \theta $  for virtual machine $i$ and back-up operating system $c$, such that $d_{ic}$ takes value one if and only if back-up operating system $c$ is assigned to virtual machine $i$. We also use the binary decision variable $e_i \ \forall i \in VM$ for virtual machine $i$ that takes value one if and only if virtual machine $i$ is assigned a back-up operating system. 

In our mathematical programming model, we penalize diversity assignments in which the same operating system is assigned to adjacent nodes. Accordingly, we use the binary decision variable $f_{ij} \ \forall (i,j) \in E$ to penalize the assigning of the same back-up operating system on endpoint $i$ and endpoint $j$ of the edge $(i,j)$.

The maximization objective function represents the expected net benefit ($ENB$), which is calculated based on Equation\ \eqref{eq:ENB}, in which $M$ represents a large enough value to be used as ``big M'' for penalizing the same back-up operating system being assigned to adjacent nodes.

\begin{equation}\label{eq:ENB}
\begin{split}
     ENB = & ALE_{\text{after}} - ALE_{\text{before}} \\
     &- \text{cost of security} - \sum_{(i,j) \in E} M  f_{ij} \\
     =  &\sum_{p \in \text{paths}} \sum_{i \in p} SLE^d_i ARO_i 
     - \sum_{p \in \text{paths}} \sum_{i \in p} SLE_i ARO_i \\
     &- \sum_{i \in VM} \sum_{c \in \theta} CS_c d_{ic}
     - \sum_{(i,j) \in E} M  f_{ij}
\end{split}
\end{equation}

The objective function in \eqref{eq:ENB} is formed first by subtracting the cost of security incurred by implementing the Diversity technique from the benefit of security that was formulated in \eqref{eq:BS}. Second, the penalty of assigning the same back-up operating system to adjacent nodes is applied to the objective function.

We formulate the term $SLE^d_i$ using the binary decision variables $d_{ic}$ and $e_i$ in \eqref{SLE-linear}. According to this linear formulation, the $SLE^d$ for virtual machine $i$ remains unchanged if no back-up operating system is assigned to it ($e_i = d_{ic} =0 \ \forall c \in \theta$). However, if back-up operating system $c$ is assigned to the virtual machine $i$ ($e_i = d_{ic} = 1$), the value of $SLE^d$ is updated based on the asset value and the exploitability factor of the assigned back-up operating system $c$.

\begin{equation}\label{SLE-linear}
\begin{split}
   SLE^d_i= SLE_i(1-e_i) + \sum_{i \in VM} \sum_{c \in \theta} d_{ic}  {AV}_c  {EF}_c 
   \end{split}
\end{equation}

The maximum expected net benefit under all possible assignments of $|\theta|$ potential back-up operating systems on $|VM|$ virtual machines is obtained by solving the {\emph{optimal diversity assignment problem (O-DAP)}} formulated in \eqref{optimization_model} as a binary linear optimization model.

\begin{equation}\label{optimization_model}
\begin{split}
&\max_{d_{ic}: i \in VM , c \in \theta, e_i: i \in VM, f_{ij}: (i,j) \in E} Z = ENB    \\
\text{s.t.} \quad
\sum_c d_{ic} &\geq e_{i} \quad \forall i \in VM \\
 f_{ij} &\geq d_{ic}+d_{jc}-1 \quad \forall (i,j) \in E, \forall c \in \theta \\
d_{ic} &\in \{0,1\} \quad \forall i \in VM , \forall c \in \theta \\
x_{i} &\in \{0,1\} \quad  \forall i \in VM \\
f_{ij} &\in \{0,1\} \quad \forall (i,j) \in E 
\end{split}
\end{equation}

Given a network of virtual machines with certain asset values and exploitability factors, before implementing any Diversity techniques (before solving the optimization problem), $ALE_{\text{before}}$ can be computed by summing $SLE \times ARO$ over all virtual machines in all attack paths. Based on the binary decision variables $d_{ic}$ and $e_i$, $ALE_{\text{after}}$ can be computed after updating the $SLE^d$ values according to \eqref{SLE-linear}.

The dependencies between the $d_{ic}$ and $e_{i}$ values are taken into account using the first constraint in \eqref{eq:ENB} (one linear constraint for each virtual machine) which also supports the natural constraint that each virtual machine gets at most one back-up operating system. The second constraint in \eqref{eq:ENB} is for obtaining the edges $(i,j)$ for which the same back-up operating system is assigned on adjacent nodes. Accordingly,
{\emph{O-DAP} model} has $(|\theta|+1)|VM|+|E|$ binary decision variables and $|VM|+|E|$ constraints. 

Generally, the problem of assigning $|\theta|$ potential back-up operating systems on $|VM|$ virtual machines has ${(|\theta|+1)}^{|VM|}$ solutions because each virtual machine can independent of others get either one of the back-up operating systems or none. Therefore, the solution to {\emph{O-DAP}} for instances with a large $|VM|$ and $|\theta|$ cannot be found by exhaustively going through all ${(|\theta|+1)}^{|VM|}$ possibilities and finding the solution with the maximum desired output. However, our formulation provided in \eqref{optimization_model} is a binary linear programming model, and can be used to efficiently find the globally optimal solutions to instances with thousands of virtual machines in a reasonable time.

\subsection{Numerical Experiment of Optimization Model}\label{ss:numerical}
In this section, we discuss a numerical example with seven potential back-up operating systems to be implemented as the Diversity technique on nine virtual machines in the upper layer of the HARM shown in Figure \ref{fig:HARM1} and solve it using \textit{Gurobi} solver. 

Table~\ref{Backup} shows an e-Health cloud equipped with various back-up OS variants that can be used for Diversity techniques. The table represents the number of patched or mitigated vulnerabilities and the cost of security for each entry, as well as the asset value for each VM. It is assumed that more secure back-up variants have higher cost of security values, and that accordingly, the damage has less impact.
\begin{table}[t]
	\centering
	\caption{back-up OS variants used for the optimization test case}
	\label{Backup}
	\begin{tabular}{|l|l|l|l|l|l|}
		\hline
		\multirow{2}{*}{No.} & \multirow{2}{*}{back-up OS ($\theta$)} & \multicolumn{2}{l|}{Vulnerabilities (V)} & \multirow{2}{*}{CS ($\$$)} & \multirow{2}{*}{AV ($\$$)} \\ \cline{3-4}
		&                                       & $|V|$             & EF               &                     &                     \\ \hline
		1                    & HP-UX 11i                             & 4                 & 0.55             & 55                  & 450                 \\ \hline
		2                    & Windows (Win 8)                       & 4                 & 0.53             & 65                  & 490                 \\ \hline
		3                    & Solaris                               & 3                 & 0.51             & 80                  & 550                 \\ \hline
		4                    & Win XP                                & 3                 & 0.49             & 100                 & 590                 \\ \hline
		5                    & CentOS                       & 2                 & 0.47             & 120                 & 620                 \\ \hline
		6                    & OpenBSD                               & 1                 & 0.45             & 150                 & 680                 \\ \hline
		7                    & Win Server 2008                               & 1                 & 0.43             & 200                 & 690                 \\ \hline
	\end{tabular}
\end{table}

Table~\ref{Backup} shows that the seven back-up systems have the costs of security, the exploitability factors, and the asset values displayed in Table \ref{tab7}. {We have selected the economic parameters consistent with our modeling assumptions for the purposes of designing an experiment to demonstrate how mathematical optimization can be used to solve the optimal diversity assignment problem. However, the values for the vulnerabilities are real metrics drawn from the National Vulnerability Database (NVD) }\cite{mell2006common}.

\begin{table}[]
\caption{Parameters computed for the seven back-up OS variants}
\label{tab7}
\begin{tabular}{|l|l|l|l|}
\hline
No. & Cost of security & Exploitability factor & Asset Value \\ \hline
1   & 55               & 0.55                  & 450         \\ \hline
2   & 65               & 0.53                  & 490         \\ \hline
3   & 80               & 0.51                  & 550         \\ \hline
4   & 100              & 0.49                  & 590         \\ \hline
5   & 120              & 0.47                  & 620         \\ \hline
6   & 150              & 0.45                  & 680         \\ \hline
7   & 200              & 0.43                  & 690         \\ \hline
\end{tabular}
\end{table}

Based on the upper layer of the HARM shown in Figure \ref{fig:HARM1}, the values for $ALE_{\text{before}}$, $SLE_i$, and $ARO_i$ are as follows:

\begin{equation*}
  \begin{aligned}
 ALE_{\text{before}}&=160194 \\
 SLE&=[300.0,300.0,300.0,300.0,283.2 \\
&    ,283.2,283.2,283.2,283.2,5900.0] \\
 ARO&=[1,1,1,1,1,1,1,1,1,1].    
 \end{aligned}
\end{equation*}

The Gurobi model for this instance of the \emph{O-DAP} is provided in the Appendix, in which the value of ``big M'' is considered to be $100,000$. 

While this instance of the problem has ${(|\theta|+1)}^{|VM|}=8^9=134217728$ feasible solutions, the Gurobi solver obtains the globally optimal solution (associated with the maximum value of the expected net benefit) in 0.02 seconds on an ordinary laptop with 8.00 GBs of RAM and Intel Core i5 6360U CPU @ 2.00 GHz.

\begin{figure}[t]
	\centering
	\begin{tikzpicture}[scale=0.79, every node/.style={transform shape}]
	\node[shape=circle,draw=black,align=center,fill=red!10]  (A) at (8.6,0) {$A$};
	\node[shape=circle,draw=black,align=center] (v1) at (7,2) {$\text{vm}_1$};
	\node[shape=rectangle,line width=0.1pt, rounded corners,opacity=.8,draw=black,align=center,fill=green!20] (W) at (7+0.45,2+0.50) {\footnotesize$W$};
	\node[shape=circle,draw=black,align=center] (v2) at (7.7,-2) {$\text{vm}_2$};
	\node[shape=rectangle,line width=0.1pt, rounded corners,opacity=.8,draw=black,align=center,fill=green!20] (W) at (7.7+0.45,-2-0.50) {\footnotesize$W$};
	\node[shape=circle,draw=black,align=center] (v4) at (6,0) {$\text{vm}_4$};
	\node[shape=rectangle,line width=0.1pt, rounded corners,opacity=.8,draw=black,align=center,fill=green!20] (W) at (6+0.65,0) {\footnotesize$W$};
	\node[shape=circle,draw=black,align=center,fill=yellow!10] (v5) at (5.5,-2.3) {$\text{vm}_5$};
	\node[shape=rectangle,line width=0.1pt, rounded corners,opacity=.8,draw=black,align=center,fill=yellow!55] (W) at (5.5+0.45,-2.3+0.50) {\footnotesize$B6$};
	\node[shape=circle,draw=black,align=center] (v3) at (4.4,2.4) {$\text{vm}_3$};
	\node[shape=rectangle,line width=0.1pt, rounded corners,opacity=.9,draw=black,align=center,fill=green!20] (W) at (4.4+0.4,2.4+0.45) {\footnotesize$W$};
	\node[shape=circle,draw=black,align=center] (v7) at (4.2,-1.1) {$\text{vm}_7$};
	\node[shape=rectangle,line width=0.1pt, rounded corners,opacity=.9,draw=black,align=center,fill=orange!25] (W) at (4.2+0.4,-1.1+0.45) {\footnotesize$L$};
	\node[shape=circle,draw=black,align=center,fill=yellow!10] (v6) at (2.8,1.1) {$\text{vm}_6$};
	\node[shape=rectangle,line width=0.1pt, rounded corners,opacity=.9,draw=black,align=center,fill=yellow!55] (W) at (2.8-0.4,1.1+0.45) {\footnotesize$B6$};
	\node[shape=circle,draw=black,align=center,fill=blue!10] (v9) at (2.1,-1.8) {$\text{vm}_9$};
	\node[shape=rectangle,line width=0.1pt, rounded corners,opacity=.9,draw=black,align=center,fill=blue!25] (W) at (2.1+0.4,-1.8-0.45) {\footnotesize$B5$};
	\node[shape=circle,draw=black,align=center] (v8) at (0.85,1.8) {$\text{vm}_8$};
	\node[shape=rectangle,line width=0.1pt, rounded corners,opacity=.9,draw=black,align=center,fill=orange!25] (W) at (0.85+0.4,1.8+0.45) {\footnotesize$L$};
	\node[shape=circle,draw=black,align=center,fill=red!10] (v10) at (-0.5,-0.2) {$\text{vm}_{10}$};

	\path [->] (A) edge node[left] {} (v1);
	\path [->] (A) edge node[left] {} (v2);
	\path [->] (v1) edge node[left] {} (v3);
	\path [->] (v1) edge node[left] {} (v4);
	
	\path [->] (v2) edge node[left] {} (v4);
	\path [->] (v2) edge node[left] {} (v5);
	
	\path [->] (v3) edge node[left] {} (v5);
	\path [->] (v3) edge node[left] {} (v6);
	
	\path [->] (v4) edge node[left] {} (v5);
	\path [->] (v4) edge node[left] {} (v6);
	
	\path [->] (v5) edge node[left] {} (v7);
	\path [->] (v5) edge node[left] {} (v9);
	
	\path [->] (v6) edge node[left] {} (v8);
	\path [->] (v6) edge node[left] {} (v9);
	
	\path [->] (v7) edge node[left] {} (v6);
	\path [->] (v7) edge node[left] {} (v9);
	
	\path [->] (v8) edge node[left] {} (v10);	
	\path [->] (v9) edge node[left] {} (v10);
	
	\end{tikzpicture}
	\caption{Optimal OS diversity assignment satisfying the coloring requirement on adjacent nodes and maximizing the expected net benefit (note that the back-up denoted by $B5$ and $B6$ are CentOS and OpenBSD, respectively.)} 
	\label{fig:final-D}
\end{figure}
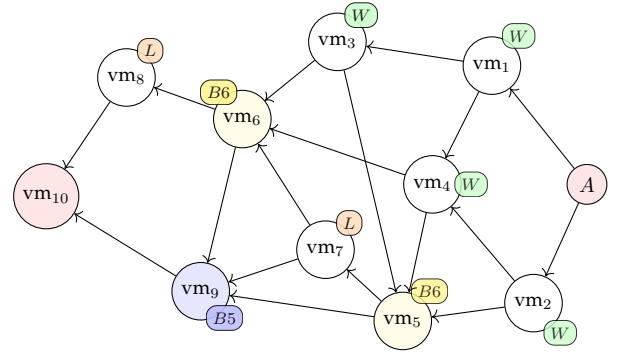

The optimal value of the expected net benefit is 117.8 which is achieved by assigning back-up operating system 6 to virtual machines 5 and 6 and back-up operating system 5 to virtual machine 9 ($d_{5,6}=1.0, d_{6,6}=1.0, d_{9,5}=1.0$). These optimal changes to the upper layer of the HARM are represented in Figure~\ref{fig:final-D}.



\section{Discussion and Limitations}\label{DL}


In Subsection~\ref{diversity}, we showed that deploying the Diversity technique can enhance the security of the cloud by increasing the $AC$ values (i.e., the attacker must expend more effort to penetrate the cloud and exploit a target). We also observed that deploying the Diversity technique decreases the $RoA$ values, which indicates that the attacker would be less likely to attack. Furthermore, we demonstrated that increasing the number of variants of the OS diversification technique leads to higher $AC$ values, and provides lower values for the $RoA$ metrics.

By comparing three strategies for deploying the Diversity technique on multiple VMs, we showed that the best OS diversification strategy is deploying the Diversity technique on VMs grouped by the BVS group. However, deploying the Diversity technique was not shown to have significant effects on the $Risk$ and $Reliability$ values. In Section~\ref{S+D+R}, we showed that combining all MTD techniques (S+D+R) could improve all aspects of security with respect to the security metrics used (i.e., the $Risk$, $AC$, $RoA$, and $Reliability$ values).

We compared D-Only to all MTD techniques combined (S+D+R), and observed that the results of deploying S+D+R were better. The values of the $RoA$ metric in S+D+R were less than those in D-Only. Moreover, the deployment of S+D+R led to more promising results for $Reliability$; see Figure~\ref{fig:combined}. Indeed, we found that deploying S+D+R could help cloud providers keep the security level of the cloud at the desired level. Figure~\ref{fig:lines} shows the overall results of combining the MTD techniques. Based on a comparison of all security metrics before and after deploying the MTD techniques, it shows the effectiveness of the deployment of MTD techniques in one scope. When we compare the metrics displayed in Figure~\ref{fig:lines}, we can see that deploying S+D+R decreased the $RoA$ and the overall risk of the cloud, while the $AC$ and $Reliability$ values increased. However, we could also change the parameters used to obtain the desired results based on the type of cloud and the required security levels. For instance, we experimented with setting the attack rate at $0.2$, and setting the OS diversification variants between $1v$ and $5v$. {We summarized the advantages and disadvantages of using single a MTD technique and the combination of the MTD techniques in Table~{\ref{comp}} based on the results obtained in this study, and on insights from the MTD survey presented in~}\cite{cho2020toward}. 

	\begin{table}[t]
	\scriptsize
	\caption{{Comparison of MTD techniques and combinations}}
	\label{comp}
	\centering
	\begin{tabular}{|p{1.25cm}|p{3.47cm}|p{2.85cm}|}
	\hline
	Type       & Pros                                                                                                                                            & Cons                                                                                            \\ \hline
	Shuffle    & \begin{tabular}[c]{@{}l@{}}Use of existing component\\ Affordable\\ Confuse the attackers \\ Increase attack difficulties\end{tabular}                                   & \begin{tabular}[c]{@{}l@{}} Inherent vulnerabilities\\ Service Interruption\end{tabular}      \\ \hline
	Diversity  & \begin{tabular}[c]{@{}l@{}} May vary existing vulnerabilities\\ Increase costs of attack\\ Waste Attacker's information\end{tabular} & \begin{tabular}[c]{@{}l@{}}Additional costs\\ Rely on available variants\end{tabular}     \\ \hline
	Redundancy & Provide reliability/availability                                                                                                              & \begin{tabular}[c]{@{}l@{}}Additional costs\\ May increase attack surface\end{tabular}    \\ \hline    
				{Combination}   & {Benefit from advantages of all}                                                                                                                & \begin{tabular}[c]{@{}l@{}}{Higher complexity}\\ {Additional costs}\end{tabular} \\ \hline
		\end{tabular}
	\end{table}

In addition, we conducted our experiment by modeling a PHC cloud example to evaluate the security and economic metrics for both the Shuffle and the Diversity MTD techniques. We leveraged a VM placement strategy for deploying the Shuffle technique. We also utilized two strategies for deploying the Diversity technique in which a single back-up OS could be deployed on either one selected VM or a multiple set of selected VMs in the cloud. We observed that deploying the Diversity technique using the same OS (one OS variant) on multiple VMs could enhance $RoSI$ values. We further found that compared to the other Diversity technique strategies, deploying four OS diversification variants among four VMs had the highest betweeness values, which led to the highest $RoSI$ values, while incurring reasonable $CS$ values.

Moreover, we solved an optimal diversity assignment problem in order to find the most promising results based on the given network and a set of various back-up OS variants using the theoretical and mathematical optimization model in Section~\ref{optimal}.
We showcased our proposed Diversity techniques using a test case of seven OS variants (back-up OS) and a cloud network with nine VMs based on the parameters shown in Table~\ref{Backup}. We used the Gurobi solver to find the optimal assignment of OS variants to virtual machines by solving a binary linear programming model (\emph{O-DAP}). In this case, the globally optimal solution among more than $134$ million possibilities was that assigning certain back-up OSs to certain VMs yielded the maximum expected net benefits as in Section~\ref{optimal}.

\vspace{1mm}
\noindent {\textbf{Limitations:}}
In this paper, we did not consider the economic metrics for combining all three MTD techniques. For each MTD technique, the security and economic metrics may be expected to vary in different ways. A multi-objective optimization was needed to solve the problem in such a way that the three MTD techniques could be combined effectively to satisfy the security and economic demands of the cloud provider based on certain constraints, such as the given model and the allocated budget.

{Moreover, we only considered OS-level vulnerabilities on each VM in order to perform the security analysis, even though there were other vulnerabilities on each cloud's components, such as network, application, and service vulnerabilities. An investigation that weighs the optimal economic investment against the vulnerability of information is presented in }\cite{gordon2002economics}{. However, based on~}\cite{gordon2002economics}{, it is important to construct a model that is able to specifically evaluate how different types of vulnerabilities, and the potential losses from such vulnerabilities, may affect the optimal security investment when combining MTD techniques.}

Finally, we deployed an optimization model for the Diversity technique, while modeling and formulating optimization models for other MTD techniques, such as the Shuffle and
Diversity techniques. Moreover, multi-objective optimization can be utilized to find an optimal solution based on several objective functions, such as RoSI and RoA. We plan to address the use of an integrated optimization model in future work. In addition, the optimal values we obtained were based on a single operation only, even though MTD techniques can be adapted and deployed on a periodic basis. While defensive MTD techniques can be deployed periodically or occasionally based on various factors, such as the annual rates of attack (ARO) or of intrusion detection, such approaches are outside the scope of this paper. However, we aim to address these issues in our future research.

\section{Conclusions}\label{Conclude}
Several techniques for enhancing the security of the cloud have been proposed. Among them, MTD strategies for mitigating potential cybersecurity threats on the cloud represent a new paradigm, and have been systematically examined over the past couple of years. However, there is a lack of research on the effectiveness of combining these techniques in order to enhance the security of the cloud. In this study, we reviewed the current state-of-the-art MTD techniques that are applicable in the cloud. Then, we used a formal security model to evaluate the effects of combined MTD techniques. In comprehensive experimental analyses, we demonstrated that a combined approach was more effective than a single MTD technique. Our proposed approach can be used for evaluating the effectiveness of individual or combined MTD techniques based on both security and economic metrics. Moreover, we evaluated the effectiveness of MTD techniques for a specific e-health cloud model to compare the costs of security and the value of the security provided. We proposed an optimization model for diversity allocation that generates the assignment associated with the maximum expected net benefit. We also showed that our binary linear programming formulation could handle large cloud models in a fraction of a second on an ordinary computer.


\bibliographystyle{splncs03}
\bibliography{MTD}

\vskip -3\baselineskip
\begin{IEEEbiography}[{\includegraphics[width=1in,height=1.25in,clip,keepaspectratio]{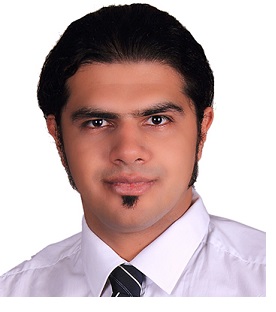}}]{Hooman Alavizadeh}
 is a Postdoc Fellow at the School of Natural and Computational Sciences, Massey University, Auckland, New Zealand. He received his PhD degree in cybersecurity from Massey University, New Zealand, and received his MSc degree in Computer Science from Eastern Mediterranean University (EMU), Cyprus. His research interests are in Moving Target Defense (MTD), cloud security, security modeling and analysis, and AI-based cyber defense.
\end{IEEEbiography}

\vskip -3.5\baselineskip
\begin{IEEEbiography}[{\includegraphics[width=1in,height=1.25in,clip,keepaspectratio]{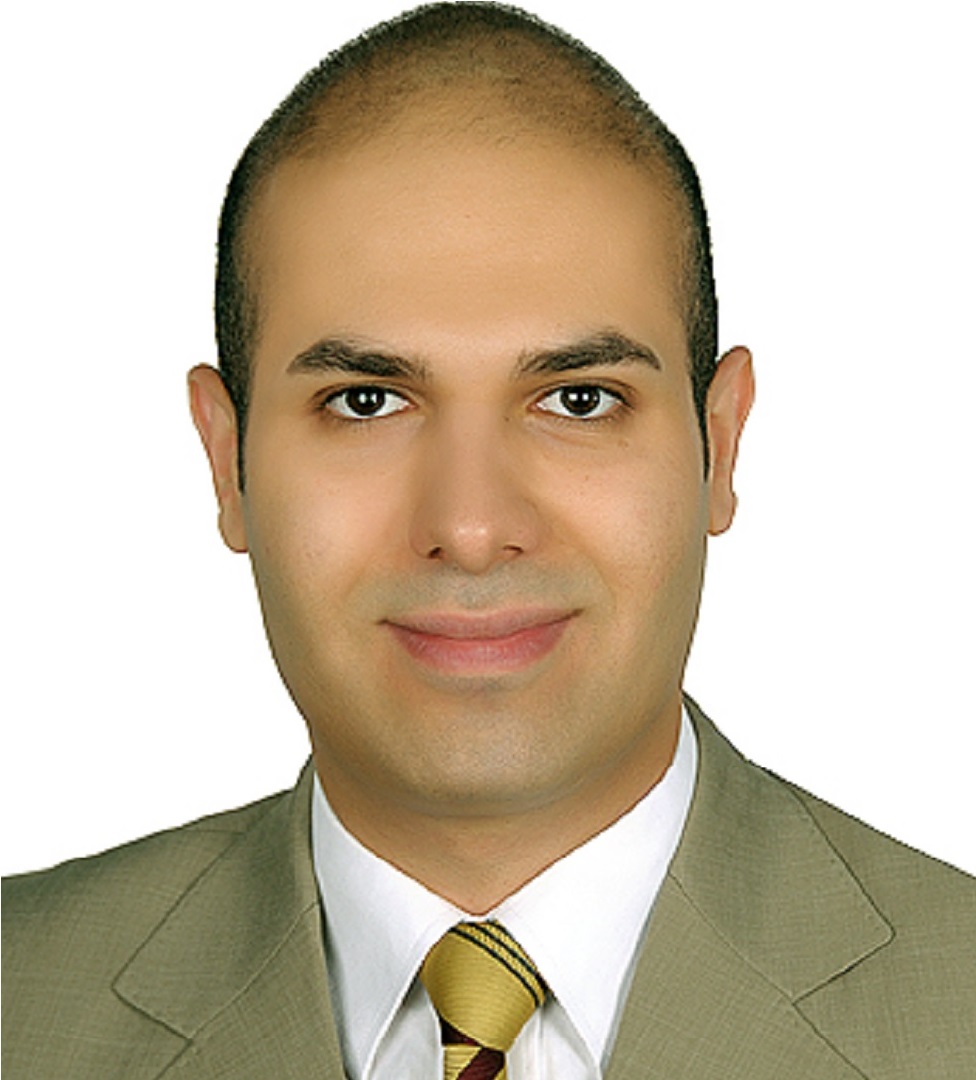}}]{Samin Aref}
	works at the Max Planck Institute for Demographic Research in Rostock, Germany. He holds a PhD in Computer Science from the University of Auckland and a MSc in Industrial Engineering and Operations Research from Sharif University of Technology. 
\end{IEEEbiography}

\vskip 4\baselineskip
\begin{IEEEbiography}[{\includegraphics[width=1.2in,height=1.25in,clip,keepaspectratio]{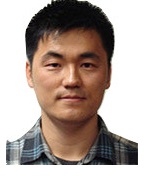}}]{Dong Seong Kim}
	is an Associate Professor in the School of Information Technology and Electrical Engineering, The University of Queensland (UQ), Brisbane, Australia. Prior to UQ, he led the Cybersecurity Lab at the University of Canterbury (UC), Christchurch, New Zealand from August 2011 to January 2019. His research interests are in cybersecurity and dependability for various systems and networks.
\end{IEEEbiography}

\vskip  -10.5\baselineskip
\begin{IEEEbiography}[{\includegraphics[width=1in,height=1.5in,clip,keepaspectratio]{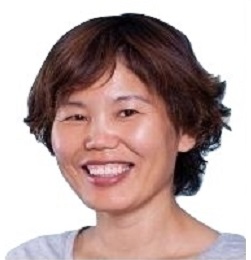}}]{Julian Jang-Jaccard}
	is an Associate	Professor in School of Natural and Computational Sciences, Massey University, Auckland, New Zealand. Julian received her PhD in database transaction (University of Sydney); a Master of Information Engineering (University of Sydney) and a Bachelor of Computer Science (University of Western Sydney). Her research interests spans from database, cloud computing, mobile systems, cybersecurity, and big data analytics with a specific focus on security and privacy. 
\end{IEEEbiography}

\section{Appendix*}

Gurobi model for the instance of the \emph{O-DAP} problem used as a test case in Subsection \ref{ss:numerical} is provided below. Properties of the seven potential back-up operating systems were provided in Table \ref{Backup}. In this instance, Diversity technique was to be implemented on nine virtual machines in the network shown in Figure \ref{fig:HARM1}.

Maximize\\
  (-269.2 \\
  + 2915.0 d1,1\\ 
  + 3051.4 d1,2\\ 
  + ... \\ 
  + 4250.5 d9,7 \\
  + -3600.0 e1 \\
  + -3000.0 e2 \\
  + ... \\
  + -4248.0 e9 \\
  + -100000.0 f1,3 \\
  + -100000.0 f1,4 \\
  + ... \\
  + -100000.0 f7,9)\\
  \\
Subject To\\
  -1.0 d1,1 + -1.0 d3,1 + f1,3 $\geq$ -1.0\\
  -1.0 d1,1 + -1.0 d4,1 + f1,4 $\geq$ -1.0\\
  ...\\
  -1.0 d7,7 + -1.0 d9,7 + f7,9 $\geq$ -1.0\\
  d1,1 + d1,2 + d1,3 + d1,4 + d1,5 + d1,6 + d1,7  $\leq$ e1\\
  d2,1 + d2,2 + d2,3 + d2,4 + d2,5 + d2,6 + d2,7  $\leq$ e2\\
  ...\\
  d9,1 + d9,2 + d9,3 + d9,4 + d9,5 + d9,6 + d9,7  $\leq$  e9\\
  \\
Binaries\\
  d1,0, d1,1, ... , d9,7, e1, e2, ... , e9, f1,3, f1,4, ... , f7,9

\end{document}